\documentclass[twocolumn,prb,floatfix]{revtex4}

\usepackage{lscape}
\usepackage{epstopdf}
\usepackage{floatrow}
\usepackage{amsmath}
\usepackage{graphicx}
\usepackage{rotating}
\usepackage{bm}
\usepackage{ulem}
\usepackage{amssymb}
\date{\today}
\usepackage{sidecap}
\usepackage[utf8]{inputenc}
\usepackage{color}
\definecolor{Blue}{rgb}{0,0,1}
\definecolor{Red}{rgb}{1,0,0}
\definecolor{Green}{rgb}{0,0.52,0.0}
\definecolor{orange}{rgb}{1,0.5,0}
\definecolor{orange2}{rgb}{1,0.5,0.5}

\newcommand{\lbl}{\label}
\begin{document}

\title{\ Finite temperature electronic structure of Diamond and Silicon}

\author{Vaishali Shah\footnote{~vaishali@unipune.ac.in}}
\affiliation{Interdisciplinary School of Scientific Computing, Savitribai Phule Pune University, Pune, 411007, India}
\author{Bhavik Sanghavi, Rahul Ramchandani, M. P. Gururajan and T. R. S. Prasanna\footnote{Corresponding author: prasanna@iitb.ac.in}}
\affiliation{Department of Metallurgical Engineering and Materials Science, Indian Institute of Technology, Bombay, Powai, Mumbai - 400076, India}

\begin{abstract}
The electron-phonon interaction contribution to the electronic 
energies is included in density functional total energy calculations 
with \textit{ab initio} pseudopotentials via 
the formalism of Allen [Phys. Rev. B, 18 5217 (1978)] to obtain the
temperature dependent electronic structure of diamond and silicon. 
This method allows us to obtain the thermally-averaged \textit{ab initio} 
electronic structure in a straightforward and computationally 
inexpensive way. Our investigations on the finite temperature electronic 
structure of diamond and silicon bring out that a new criterion, that of 
temperature transferability, is required in the input 
\textit{ab initio} pseudopotentials for temperature dependent studies. 
The temperature transferability of the Troullier-Martins pseudopotentials 
used in this work is strongly dependent on the cut-off radius and the 
inclusion of the unbound 3d$^0$ state. The finite temperature indirect
band gaps are highly sensitive to the choice of cut-off radius used
in the pseudopotentials. The finite temperature band structures
and density of states show that thermal vibrations affect the electron
energies throughout the valence and conduction band. We compare our results on the 
band gap shifts with that due to the Debye-Waller term in the Allen-Heine 
theory and discuss the observed differences in the zero point and high temperature band gap shifts. 
Although, the electron energy shifts in the highest occupied valence band and lowest 
unoccupied conduction band enable to obtain the changes in the 
indirect and direct band gaps at finite temperatures, the shifts in
other electronic levels with temperature enable investigations into 
the finite temperature valence charge distribution in the bonding region. 
Thus, we demonstrate that the Allen theory provides a simple and theoretically 
justified formalism to obtain finite temperature valence electron charge 
densities that go beyond the rigid pseudo-atom approximation.

\end{abstract}

\maketitle
\section{Introduction}

Most electronic structure studies are performed for static lattices that 
are implicitly assumed to be at 0 K. Experimental investigations and 
determination of material properties are however performed at finite 
temperatures. The two important effects of temperature on the material 
are the lattice expansion and lattice dynamics. 
The effect of the lattice expansion on electronic energies is 
straightforward to obtain. However, the electron-phonon interaction 
which has a major contribution from the lattice dynamical behaviour is 
harder to calculate. Temperature affects the nuclear motion in materials 
leading to lattice dynamics which alters the electronic energies by 
2-4 k$_{B}$T~\cite{AllenHeine,Allen,AllenHui} in solids.
The resistivity of metals, directional Compton profiles, infrared,
Raman, optical spectra, specific heats, heat conduction, band gaps etc. 
are affected by the electron-phonon interaction.
In order to compare the results of experiments with the theoretical 
simulations the effect of electron-phonon interaction needs to be 
included in the calculations of these phenomena.~\cite{Allen, AllenHui, Giustino2017}.

Although of significant importance, the electron-phonon interaction 
happens to be the most difficult to compute from first principles. 
Currently, there are three major approaches~\cite{PoncePRB2014}, namely, 
the molecular dynamics method, the frozen phonon method and the 
perturbation theory method to understand 
the effects of the electron-phonon interaction on material properties. 
Each of these approaches has its advantages and 
disadvantages~\cite{PoncePRB2014, Giustino2017}. The to-date developments 
to include the electron-phonon interactions in \textit{ab initio} 
calculations and their computational implementations is discussed in 
detail in a rigorous and elegant review by Giustino~\cite{Giustino2017}.

In the context of temperature dependent semiconductor band gaps,
the earliest attempt to include the electron-phonon interaction 
was by Fan~\cite{Fan} who calculated the self-energy (SE) contribution to 
the semiconductor band gap shifts. This term is also referred to as the Midgal term
in superconductor literature~\cite{Giustino2017}. Soon afterwards, Antoncik~\cite{Antoncik}
calculated the Debye-Waller (DW) correction to the semiconductor band gaps.  
Subsequently, several authors calculated the semiconductor band gap shifts 
due to thermal vibrations, mostly by incorporating the Debye-Waller term~\cite{Keffer, WalterCohen, Kasowski, Schluter, Cardona2005}. 
These investigations led Baumann~\cite{Baumann} to suggest that 
both self-energy and DW corrections are necessary for a complete account 
of the role of thermal vibrations on electron energies. 
 
Allen and Heine~\cite{AllenHeine} developed the formal basis for 
incorporating electron-phonon interactions in the harmonic approximation at 
constant volume using second-order perturbation theory. In this 
formalism~\cite{AllenHeine},  the DW term and the SE term appear separately 
and the total electron energy shift is a sum of the two terms.
The implementation of the formalism requires a concurrent 
calculation of the DW and SE term. 
The calculation of the DW term is approximated by a second 
order expansion with the neglect of the higher order terms. 
Further, by exploiting translational invariance the DW term 
(in the second order form) and the SE term is recast~\cite{AllenHeine} to 
have similar forms.  All recent studies based on the Allen-Heine theory use 
the recasted DW term since calculating it directly in the second order 
form is computationally complex.

The Allen-Heine theory was earlier used with empirical pseudopotentials to 
obtain the band gap shifts in semiconductors~\cite{AllenCardona, Zollner}. 
However, in the last decade, several \textit{ab initio} studies~\cite{Marini, 
GiusCohen2010, PonceCMSc2014, Gonze2014, PatrickGiustino2014, ZachGius2016, Ponce2015, MonserratNeeds2013, MonserratNeeds2014} 
based on it have been reported. These studies report the band gap shifts at specific k points that 
usually correspond to the indirect and direct band gaps~\cite{Marini, 
GiusCohen2010, PonceCMSc2014,Gonze2014, PatrickGiustino2014, ZachGius2016, Ponce2015, 
MonserratNeeds2013, MonserratNeeds2014}. 
In these studies, the band gaps at all high symmetry points or the complete
valence electron band structure along the symmetry lines in the Brillouin 
zone have not been reported. 

Subsequent to the Allen-Heine theory~\cite{AllenHeine}, a more 
accurate theory was developed by Allen~\cite{Allen} for the temperature 
dependent band structure. In this theory, the DW term is 
considered to all orders instead of its truncated second 
order expansion as in the Allen-Heine theory. In the implementation
of the Allen theory~\cite{Allen} with pseudopotentials, the first step is to 
correct the pseudopotential form factor with the DW term and use this 
corrected pseudopotential to obtain the finite temperature electronic 
structure. The second step is to calculate the SE term using the wavefunctions 
and energies obtained from the DW corrected electronic structure. 

The Allen theory~\cite{Allen} provided the theoretical justification for the 
earlier empirical studies~\cite{Keffer, WalterCohen, Kasowski, Schluter, 
Cardona2005} that calculated the finite temperature band gaps
based on the DW correction step. It must be noted that the first step, the 
DW correction step, directly leads to the temperature dependent 
band structure. The band gap shifts obtained from the first
DW correction step gave satisfactory band gap shifts in some cases and unsatisfactory results in
other cases~\cite{Keffer, WalterCohen, Kasowski, Schluter, 
Cardona2005}. The DW corrected empirical studies on the temperature dependence of
band gaps in PbSe and PbTe, while giving reasonable results for the direct
gap, failed to correctly estimate the indirect gap\cite{Schluter}. 
Allen and Heine~\cite{AllenHeine} and Allen~\cite{Allen} discuss this aspect 
and show that for accurate values of the finite temperature electron energies and
 band gap shifts the contribution from the SE term needs to be included, in general.

The second order Allen-Heine formalism~\cite{AllenHeine} continues 
to be of interest and has been used in several recent \textit{ab initio} 
calculation based studies~\cite{Marini, GiusCohen2010, PonceCMSc2014, Gonze2014, 
PatrickGiustino2014, ZachGius2016, Ponce2015, MonserratNeeds2013, 
MonserratNeeds2014} to obtain the semiconductor band gap shifts 
with temperature. In contrast, the Allen theory~\cite{Allen} has not been 
used in \textit{ab initio} studies, even though it is a more accurate 
approach~\cite{AllenHeine} compared to the Allen-Heine theory. 

In this article we demonstrate that the 
Allen theory can be combined with \textit{ab initio} pseudopotentials 
to obtain \textit{ab initio} finite temperature electronic structures 
without any additional increase in the computational complexity of these 
calculations. 
We investigate the finite temperature electronic structure of 
diamond and silicon using the \textit{ab initio} pseudopotential implementation 
of the Allen theory. We then compare the results 
of our theoretical calculations with earlier \textit{ab initio} studies on 
diamond and silicon based on the second-order Allen-Heine theory.

The focus of this paper is to address the main issues that affect the  
implementation of the Allen theory~\cite{Allen} using \textit{ab initio} 
pseudopotentials. Thus, only the DW contribution, which is the first step 
in Allen theory, has been calculated. The (Fan/Fan-Migdal)
self-energy term, that is to be calculated from the results of the DW 
step, has not been evaluated.

However, calculation of the DW correction to the electronic energies 
is important for three reasons, as will be seen in this study.
Firstly, the DW step is sufficient to address the fundamental question, viz.
the viability of implementing Allen theory with \textit{ab initio} pseudopotentials. 
Secondly, only the DW term can be the basis for the 
comparison of the Allen and Allen-Heine theories. And, thirdly, 
the DW step is sufficient to obtain finite temperature 
valence electron charge densities.  Thus, the 
results of the present study have important implications for 
\textit{ab initio} finite temperature electronic structure studies.

In the next section, we briefly describe the theory of Allen~\cite{Allen} for the 
inclusion of the electron-phonon interaction and its implementation within 
the pseudopotential method for band structure calculations. 
In Sec. \ref{sec-com}, we describe the computational implementation and 
give the details of our calculations of temperature dependent electronic 
structure. 
Sec. \ref{sec-rd} discusses the results of the temperature dependent band 
gap trends that were obtained using existing pseudopotentials and  
the need to use temperature transferable pseudopotentials in 
electronic structure calculations that are to be used to study the 
finite temperature properties. 
Our results on the temperature dependent band gap trends, band gap 
shifts at zero point vibration and higher temperatures for diamond 
(Sec. \ref{tbs-c}) and silicon (Sec. \ref{si}) are discussed in detail 
with implications.
In the Sec. \ref{ft-vcd}, we show that theoretically it is
possible to obtain charge densities that go beyond the rigid pseudo-atom 
approximation for a direct comparison with experimental data and this 
is then followed by conclusions in Sec. \ref{concl}

\section{Allen’s theory and its implementation}\lbl{Allen-Th}

In the theory of temperature dependence of the energy bands given by
Allen-Heine~\cite{AllenHeine} and Allen~\cite{Allen}, a lattice of identical 
atoms is considered with the atoms undergoing a small thermal 
displacement $\mathbf{u}_\alpha$ about their equilibrium positions 
$\mathbf{\alpha}$. In this system, the 
potential experienced by an electron due to phonon disorder is assumed 
to move rigidly with the atoms so that the perturbation to the 
system can be expressed as

\begin{equation}
H_{e-p} = \sum_l [V(\mathbf{r}-\mathbf{R}_\alpha)-V(\mathbf{r}-{\mathbf{\alpha}})]
\lbl{h-ep}
\end{equation}

where, V(r) is the atomic potential, $\mathbf{R}_\alpha$ is the displaced 
position of the atom, $\mathbf{\alpha}$ is the equilibrium position 
and $\mathbf{u}_\alpha = \mathbf{R}_\alpha - \mathbf{\alpha}$ is
the displacement. The thermal displacements of atoms are time dependent and are 
related to the phonon frequency. The perturbed electron energy can be calculated in the 
adiabatic approximation by a second order Taylor expansion of 
Eq.~\ref{h-ep} as proposed by Allen-Heine~\cite{AllenHeine}, where the 
first two terms are 

\begin{equation}
H_{e-p}^{(1)} = \sum_l \Big[\frac{{\partial V}(r-\alpha)}{{\partial r}_{n}} \Big]u_{\alpha n}
\end{equation} 

\begin{equation}
H_{e-p}^{(2)} = \dfrac{1}{2} \sum_l\Big[\frac{{\partial^2 V}(r-\alpha)}{{\partial r}_{n} {\partial r}_{m}}\Big] u_{\alpha n} u_{\alpha m}
\lbl{set}
\end{equation}

Considering the thermal average $ \langle H_{e-p} \rangle $, the only non-vanishing terms
are the even powered terms like Eq.~\ref{set} in the Taylor expansion. The thermal average
of Eq.~\ref{set} is the self energy correction in a Bloch-wave basis set. However, in a plane wave
basis set all the non-vanishing terms can be considered and result in the Debye-Waller corrections 
to the crystal potential and Eq.~\ref{h-ep} can be rewritten as

\begin{equation}
H_{e-p} = \sum_{\mathbf{k}\mathbf{k}^\prime} V(\mathbf{k}^\prime - \mathbf{k})s(\mathbf{k}^\prime - \mathbf{k})c_{\mathbf{k}^\prime}^{\dagger}c_{\mathbf{k}}
\end{equation}

In this equation the $\mathbf{k}$, $\mathbf{k^\prime}$ 
span all the Brillouin zones and s is the structure factor. 
The thermal average of the perturbed Hamiltonian in reciprocal space is 
\begin{equation}
\langle H_{e-p} \rangle = \sum_{\mathbf{k}}^{BZ} \sum_{\mathbf{G}\mathbf{G}^\prime} V(\mathbf{G}^\prime - \mathbf{G})(e^{-W(\mathbf{G}^\prime - \mathbf{G})}-1)c_{\mathbf{k}+\mathbf{G}^\prime}^+ ~c_{\mathbf{k}+\mathbf{G}}
\lbl{avg-h-ep}
\end{equation}

This Hamiltonian is periodic and the calculation in Fourier space requires the
sum to be performed only on the first Brillouin zone. When added to the 
unperturbed Hamiltonian it leads to a reduction of the pseudopotential form 
factors V(G) of the static lattice by DW factors $e^{-W(G)}$ to 
incorporate the effect of the DW term of the electron-phonon interaction.
This result of Allen~\cite{Allen} provided the theoretical basis for
the earlier empirical studies~\cite{Keffer, WalterCohen, Kasowski, Schluter, Cardona2005} and
 is also the theoretical basis for the present \textit{ab initio} study. 

The underlying assumptions in both the Allen-Heine~\cite{AllenHeine} and Allen~\cite{Allen}
theories are the adiabatic approximation and the rigid atom approximation. 
The \textit{ab initio} pseudopotentials are 
generated under the assumption of frozen-core approximation for the core 
electrons, which satisfies the rigid-atom approximation.  
Thus, Allen-Heine~\cite{AllenHeine} and Allen~\cite{Allen} theories are valid 
for semiconductors and insulators at all temperatures. 
The adiabatic approximation condition, however, breaks down for metals at low 
temperatures since the self-energy term cannot be correctly 
represented~\cite{AllenHeine, Allen, AllenHui}. 

From Eq.~\ref{avg-h-ep}, the finite temperature pseudopotential form 
factor for any ion can be written as     

\begin{equation}
V_i^{ps}(\mathbf{G},T) = V_i^{ps}(\mathbf{G},0)e^{-W_i(\mathbf{G},T)}
\lbl{dw}
\end{equation}

where the DW factor~\cite{Keffer,Kasowski,Slater} is given by,  
$W_i = \dfrac{\langle u_i^2 \rangle |G|^2}{2}$  
and $\langle u_i^2 \rangle$ is the mean square displacement 
of atom i. 

In order to use Eq.~\ref{dw} in \textit{ab initio} studies, 
both the pseudopotential form factor, V(G) and the DW factor (W) need to be obtained from 
\textit{ab initio} studies. 
Currently, there are several \textit{ab initio} 
methods~\cite{Schowalter, Yang, Erba2013, Erba2012, Vila, Baroni, Parlinski1997, Parlinski2007} to 
calculate DW factor from first principles. 
In these studies, \textit{ab initio} DW
factor has already been calculated for several materials including many 
semiconductors. 

However, if necessary, even experimental DWF can be used in Eq.~\ref{dw}.
We note that in several \textit{ab initio} studies~\cite{Erba2013, Erba2012, Ramirez} experimental lattice 
parameters are used. In particular, Erba et. al~\cite{Erba2013} provide 
the justification for the use of experimental lattice parameters viz. that 
they can be validated by separate \textit{ab initio} studies. Therefore, if 
necessary, the same justification can be the basis for the use of 
experimental DWF in \textit{ab initio} studies based on Allen theory.

The other requirement is that of \textit{ab initio} 
pseudopotentials which can be used in Eq.~\ref{dw} to obtain 
finite temperature band structure. As discussed in Sec.\ref{sec-rd}, this is 
a much more stringent condition than expected.

It is of interest to compare both the zero-point and the higher 
temperature band gap shifts obtained using the Allen and Allen-Heine 
theories since the latter uses the second-order approximation to the DW 
term.
In both, Allen and Allen-Heine theories, the band gap shifts should 
increase with temperature due to the increase in the mean-square 
displacements or W. The main difference in their formalism is that, in 
the Allen-Heine theory, the pseudopotential form factor V$_G$ e$^{-W}$ is 
approximated~\cite{AllenHeine} by V$_G$ (1-W) neglecting higher order terms in the expansion. 
The band gaps obtained by the DW corrected pseudopotentials, V$_G$ e$^{-W}$  
in the Allen theory and V$_G$ (1-W) in the Allen-Heine theory, are to be 
compared with the band gaps obtained with the static lattice pseudopotential 
form factor, V$_G$, to obtain the finite temperature band gap shifts. 
 
We can examine the effect of the second-order approximation used in 
the Allen-Heine theory.  For higher values of the mean square displacement, 
the pseudopotential form factor in the Allen-Heine theory,
V$_G$ (1-W), is smaller than that of the actual value, V$_G$ e$^{-W}$, that 
is used in the Allen theory. This implies that the pseudopotential form 
factor used in the Allen-Heine theory undergoes a larger amount of change 
from the static lattice value (V$_G$) than the pseudopotential form factor 
in the Allen theory.
It follows that, at higher temperatures, the Allen-Heine theory should give 
larger band gaps shifts than Allen theory due to the neglect of higher order 
terms.  

In the Allen-Heine theory, the self-energy (SE) term contribution 
is calculated from the static lattice electron wavefunctions and energies. 
In contrast, in the Allen theory, the SE term contribution is to 
be calculated from the finite temperature electron wavefunctions and energies 
obtained from the DW correction step.  Thus, the SE term contributions in the Allen 
and the Allen-Heine theories will be different.

Therefore, the primary basis for the comparison of the Allen-Heine and 
Allen theories is the Debye-Waller (DW) term contribution, which, in 
principle, must give identical band gap shifts for the same pseudopotential 
and same mean-square displacement values as long as the 
second-order approximation is valid.

\section{Computational Methodology}\lbl{sec-com}

We have performed \textit{ab initio} total energy calculations based on density
functional theory to obtain the band structure of diamond and silicon at 
different temperatures using the Quantum Espresso (QE) software package~\cite{QE}.
The equilibrium lattice constant for the static lattice (0 K) was obtained by 
choosing the energy converged k-point mesh and kinetic energy cutoff. 
A 6x6x6 k-point mesh and kinetic energy cutoff of 60 Ry for carbon (diamond) 
and 40 Ry for silicon was used in all our calculations.

To incorporate the effect of finite temperature on the electronic structure,
we modified the QE code so that the local part of the \textit{ab initio}
atomic pseudopotential of the system under investigation is altered in
G-space with the Debye-Waller factor, i.e. e$^{-W}$ as in Eq.~\ref{dw}.
The non-local part of the atomic pseudopotential is unmodified to 
preserve the angular dependence of the scattering potential as in the earlier
work based on empirical pseudopotentials~\cite{Schluter}. 
With this alteration, the finite temperature pseudopotential form factor is 
obtained within the QE code for a given \textit{ab initio} static lattice
pseudopotential of any atom. 
Using a value of the mean square displacement $\langle u_i^2 \rangle$ 
appropriate for a chosen temperature, an electronic structure calculation 
was performed to obtain the band structure at that temperature. 

All electronic structure calculations were performed for various 
temperatures up to 1000 K at constant volume using the equilibrium
lattice constant obtained for the static lattice. 
The \textit{ab initio} mean-square 
displacement $\langle u_i^2 \rangle$ values of C and Si listed in 
Table~\ref{msdv} at various temperatures were taken from the
\textit{ab initio} studies of Schowalter \textit{et al.}~\cite{Schowalter}. 
Using these values in the modified QE code, the self-consistent 
total energy and the band structure at different
temperatures were obtained using different kinds of 
\textit{ab initio} pseudopotentials.

\begin{table}[h!]
\begin{tabular}{lcc} \hline \hline
 T (K)  & \multicolumn{2}{c}{$\langle u_i^2 \rangle$ (\AA$^2$) } \\
        &  C        &~~ Si  \\ \hline
 0.001  &  0.001611 &~~ 0.002471 \\
  100   &  0.001626 &~~ 0.003196 \\
  200   &  0.00169  &~~ 0.004865 \\
  300   &  0.001807 &~~ 0.006788 \\
  400   &  0.001968 &~~ 0.008772 \\
  600   &  0.002436 &~~ 0.01287  \\
  800   &  0.002962 &~~ 0.017022 \\
  1000  &  0.003529 &~~ 0.021198 \\ \hline \hline
\end{tabular}
\caption{ \textit{Ab initio} mean-square displacement values for diamond 
and silicon at various temperatures~\cite{Schowalter}.}
\lbl{msdv}
\end{table}

\section{Results and Discussion}\lbl{sec-rd}

We began our investigations on the effect of temperature on the electronic
structure of diamond and silicon with several norm conserving pseudopotentials 
that are available in the pseudopotential library on 
the QE website~\cite{QE}. In addition, some 
optimized and ultrasoft pseudopotentials available on the QE 
website library, recently developed Optimized Norm Conserving 
Vanderbilt (ONCV) pseudopotentials~\cite{ONCV}, the GBRV 
pseudopotentials~\cite{GBRV} and PAW pseudopotentials \cite{virtualvault} 
were also investigated. 

Separate electronic structure calculations were performed to obtain the band 
structure of diamond and silicon for each temperature up to 1000 K as given 
in Table~\ref{msdv} using each pseudopotential listed in Table~\ref{ppc-db} 
and~\ref{ppsi-db}.  The band gaps obtained for each pseudopotential 
were then compared across the range of temperatures used in this study. 
Table~\ref{ppc-db} and~\ref{ppsi-db} list the trends in the band gaps obtained
for diamond and silicon from our calculations. The equilibrium lattice parameter
obtained for each pseudopotential is listed along with the experimental
value for comparison.

\begin{table}[!h]
\begin{tabular}{lcllcc} 
\hline \hline
 Pseudopotential &  a  &\multicolumn{4}{c} {Band gap (eV)} \\ \cline{3-6}
 (PP) Filename &  a.u.  &\multicolumn{2}{c} {Static} & \multicolumn{2}{c}{Finite T trend} \\ 
               &        & Indirect & Direct & Indirect & Direct \\ \hline
C.pbe-mt$_{-}$gipaw     & 6.73& ~4.246 & 5.645 & Decrease & Increase  \\
C.pbe-n-kjpaw            & 6.74  & ~4.135 & 5.599 & Decrease & Increase \\
C.pbe-mt$_{-}$fhi       & 6.63& ~4.308 & 5.756 & Decrease & Increase  \\
C.pw-mt$_{-}$fhi        & 6.63& ~4.32  & 5.757 & Decrease & Increase  \\
C.blyp-mt               & 6.79& ~4.371 & 5.679 & Decrease & Increase  \\
C.blyp-hgh              & 6.79& ~4.317 & 5.664 & Decrease & Increase  \\
C.pbe-hgh               & 6.77& ~4.068 & 5.589 & Decrease & Increase  \\
C.pbe-van$_{-}$bm       & 6.64& ~4.234 & 5.722 & Decrease & Increase  \\
C$_{-}$pbe$_{-}$v1.2.uspp.F& 6.71& ~4.176 & 5.629 & Decrease & Increase \\
C$_{-}$ONCV$_{-}$PBE-1.0 & 6.54  & ~4.378 & 5.846 & Decrease & Increase \\ 
Expt$^{\cite{Ponce2015}}$& 6.74  & ~5.48   & 7.3      & Decrease & Decrease \\ \hline \hline
\end{tabular}

\caption{The indirect and direct band gap at 0 K and the trend in their values 
with increasing temperature (up to 1000 K) for the carbon (diamond) \textit{ab initio} 
pseudopotentials.  The names of the pseudopotential files are the same as in 
the source library with a .UPF extension}
\label{ppc-db}
\end{table}

\begin{table}[!h]
\begin{tabular}{lcllcc} 
\hline \hline
 Pseudopotential &  a  &\multicolumn{4}{c} {Band gap (eV)} \\ \cline{3-6}
 (PP) Filename &  a.u.  &\multicolumn{2}{c} {Static} & \multicolumn{2}{c}{Finite T trend} \\ 
               &        & Indirect & Direct & Indirect & Direct \\ \hline
Si.pbe-mt$_{-}$gipaw  &10.31& ~0.633 & 2.572 & Decrease & Decrease \\
Si.pbe-n-kjpaw        &10.33& ~0.605 & 3.122 & Increase & Decrease \\
Si.pbe-mt$_{-}$fhi    &10.33& ~0.614 & 2.561 & Increase & Increase \\
Si.pw-mt$_{-}$fhi     &10.17& ~0.457 & 2.568 & Increase & Increase \\
Si.blyp-hgh           &10.41& ~0.905 & 2.849 & Increase & Decrease \\
Si.pbe-n-van          &10.34& ~0.614 & 3.108 & Increase & Increase \\
Si.pbe-rrkj           &10.35& ~0.639 & 2.558 & Increase & Increase \\
Si$_{-}$pbe$_{-}$v1.uspp.F &10.33& ~0.691 & 2.533 &  Increase & Increase \\
Si$_{-}$oncv$_{-}$pbe-1.0  &10.37& ~0.62 & 2.548 & Increase & Dec-Inc  \\
Si$_{-}$oncv$_{-}$pbe-1.1  &10.35& ~0.60 & 2.546 &  Increase & Dec-Inc  \\
Expt$^{\cite{Ponce2015}}$  &10.26& ~1.17 & 3.378 & Decrease & Decrease \\ \hline \hline

\end{tabular}
\caption{The indirect and direct band gap at 0 K and the trend in their
values with increasing temperature (up to 1000 K) for the 
silicon \textit{ab initio} pseudopotentials. The names of the 
pseudopotential files are the same as in the source library with a 
.UPF extension}
\label{ppsi-db}
\end{table}

For the static lattice case, almost all of the above pseudopotentials show
excellent agreement in values of the indirect and direct band gaps reported in
theoretical studies as all these pseudopotentials have hitherto been 
developed for static lattice \textit{ab initio} calculations. 

Several previous studies~\cite{AllenCardona, Zollner, Marini, GiusCohen2010, PonceCMSc2014,
 Gonze2014, PatrickGiustino2014, ZachGius2016, Ponce2015, MonserratNeeds2013} on diamond and silicon, based
on the Allen-Heine theory, using both \textit{ab initio} and empirical 
pseudopotentials, have shown that the indirect and direct band gaps
decrease with temperature in agreement with experiments. In particular, 
for diamond~\cite{Zollner, GiusCohen2010, PonceCMSc2014} and silicon~\cite{AllenCardona, Zollner}, the DW term alone 
leads to a decrease in the band gaps with temperature as calculated in the Allen-Heine theory.

Thus, the primary criterion that needs to be fulfilled when 
choosing \textit{ab initio} pseudopotentials for temperature dependent
studies on carbon (diamond) and silicon is that after modification with 
the DW factor, the direct and indirect band gaps must decrease with increasing 
temperatures. Only after this primary criteria is met can further finite 
temperature studies be performed. For example, the self-energy contributions 
should only be calculated for pseudopotentials which show the correct 
behaviour in the DW step. 

We notice that, for diamond (Table~\ref{ppc-db}) all the pseudopotentials show 
decreasing indirect band gaps as is expected with increasing temperature. 
However, the direct band gaps are increasing with temperature which does not 
agree with the trends reported in literature. Thus, none of the \textit{ab initio} 
pseudopotentials in Table~\ref{ppc-db} can be 
used for finite temperature electronic structure studies. 

In the case of silicon only one pseudopotential, Si.pbe-mt$_{-}$gipaw.UPF, gave 
the expected trend of decreasing indirect and direct band gaps with increasing 
temperature. The lattice constant obtained with this pseudopotential also 
gives a better agreement with the experimental value.
The calculated band gaps and the shifts in the band gaps for this 
Si pseudopotential, labelled Si$_{pbe-gipaw}$ in our study, are listed in 
Table~\ref{si-gipaw-bg}.

\begin{table}[!h]
\begin{tabular}{lllrr} \hline \hline
Temp &  \multicolumn{2}{c}{Band gap} & \multicolumn{2}{c}{Band gap shift} \\
 (K) &  \multicolumn{2}{c} {(eV)} & \multicolumn{2}{c}{(meV)} \\
           &  Indirect  &  Direct  & Indirect & Direct  \\ \hline
0     &  0.633  &    2.572  &      &     \\ 
0.001 &  0.539  &    2.544  &  94  &  28 \\ 
100   &  0.512  &    2.537  & 121  &  35 \\ 
200   &  0.449  &    2.519  & 184  &  53 \\ 
300   &  0.376  &    2.494  & 257  &  78 \\ 
400   &  0.296  &    2.466  & 337  & 106 \\ 
600   &  0.136  &    2.396  & 497  & 176 \\ 
800   &   -     &    2.319  &  -   & 253 \\ 
1000  &   -     &    2.246  &  -   & 326 \\ 
\hline \hline
\end{tabular}
\caption{The indirect and direct band gaps with temperature and the energy 
shifts in the direct and indirect band gaps for silicon obtained with
Si.pbe-mt$_{-}$gipaw.UPF (Si$_{pbe-gipaw}$) pseudopotential}
\label{si-gipaw-bg}
\end{table}

It is well known that the Si band gaps are underestimated in the density 
functional calculations. A consequence of this underestimation is that
at higher temperatures of 800 K and 1000 K the indirect band gap of 
Si is seen to vanish. As will be discussed in detail later (Sec- \ref{si}), 
while the band gap trends are correct, the actual values of the band gap shifts
 for the above pseudopotential, Si$_{pbe-gipaw}$, appear to be excessive 
 when compared to literature values.

The finite temperature band gap results show that, in general, the 
\textit{ab initio} pseudopotential developed for static lattice applications 
cannot be directly used for temperature dependent electronic structure 
calculations.
The above results thus imply that for finite temperature studies, 
\textit{ab initio} pseudopotentials need to be generated so as to satisfy an 
additional criterion, that of temperature transferability. 

Since the \textit{ab initio} pseudopotential is initially generated for static 
lattice conditions, it has to be generated with the usual criteria viz. the norm 
conservation criteria (present or absent) and the choice of the cut-off 
radius, r$_c$, to ensure the transferability of the pseudopotential to other 
chemical environments. The additional criterion of temperature transferability 
will ensure that the pseudopotentials can be used for finite 
temperature studies, in addition to static lattice calculations.

Semiconductors are the class of materials that are most suited to test for
 temperature transferability. This is because there is a stringent test of 
temperature transferability - the indirect and direct band gaps must exhibit
the ``Varshni effect'' of redshift with increasing temperature \cite{Varshni}. 
In addition, the band gap shifts must be of the  same order of magnitude when 
compared with existing experimental or 
theoretical results.

However, in some semiconductors the band gaps exhibit a blueshift with 
temperature~\cite{Schluter,Giustino2017}. For such semiconductors, the temperature transferability criterion 
stands appropriately modified.

\subsection{Pseudopotentials with temperature transferability}\lbl{pp-tt}

Considering that several of the currently available open source C and 
Si \textit{ab initio} pseudopotentials (Table~\ref{ppc-db} and~\ref{ppsi-db}) when used for 
finite temperature electronic structure 
calculations did not reproduce the correct band gap trends and band gap shifts, 
we generated pseudopotentials based on the Troullier-Martins method~\cite{TM1991}, 
with GGA/PBE~\cite{PBE} and LDA/PZ~\cite{PZ1, PZ2} exchange-correlation functionals, using 
the OPIUM~\cite{OPIUM} pseudopotential generation code. 

Table~\ref{ppc-g} and~\ref{ppsi-g} list the details of the carbon and 
silicon pseudopotentials (PP) that were generated and tested 
for temperature transferability. The cut-off radius, the number of valence 
states and their electron occupancies were varied to study their effect on 
temperature dependent band gap behavior and to identify pseudopotentials 
that exhibit temperature transferability. \newline

\begin{table}[!h]
\begin{tabular}{llccll} \hline \hline
PP & \multicolumn{3}{c}{Parameters}&  \multicolumn{2}{c}{Band gap} \\ \cline{2-4}
Name & ~Valence   & ~r$_c$ & ~~~E$_{xc}$ & \multicolumn{2}{c}{Finite T trend} \\ \cline{5-6} 
     & ~electrons &~(a.u.) &   & Indirect & Direct \\ \hline 
C$_1$     & 2s$^2$ 2p$^2$          & 1.5  & PBE  & Decrease & Decrease \\
C$_2$     & 2s$^2$ 2p$^{1.5}$      & 1.5  & PBE  & Decrease & Decrease \\
C$_3$     & 2s$^2$ 2p$^1$          & 1.5  & PBE  & Decrease & Decrease \\
C$_4$     & 2s$^2$ 2p$^2$          & 1.5  & PZ   & Decrease & Decrease \\
C$_5$     & 2s$^2$ 2p$^2$          & 1.2  & PBE  & Increase & Decrease \\
C$_6$     & 2s$^2$ 2p$^{1.5}$      & 1.2  & PBE  & Increase & Decrease \\
C$_7$     & 2s$^2$ 2p$^1$          & 1.2  & PBE  & Increase & Decrease \\
C$_8$     & 2s$^2$ 2p$^2$ 3d$^0$   & 1.5  & PBE  & Decrease & Decrease \\
C$_9$     & 2s$^2$ 2p$^{1.5}$ 3d$^0$ & 1.5 & PBE & Decrease & Decrease \\
C$_{10}$  & 2s$^2$ 2p$^1$ 3d$^0$   & 1.5  & PBE  & Decrease & Decrease \\
C$_{11}$  & 2s$^2$ 2p$^2$ 3d$^0$   & 1.5  & PZ   & Decrease & Decrease \\ 
C$_{12}$     & 2s$^2$ 2p$^2$ 3d$^0$   & 1.3  & PBE & Inc-Dec & Decrease \\
C$_{13}$     & 2s$^2$ 2p$^2$ 3d$^0$   & 1.4  & PBE & Inc-Dec & Decrease \\
C$_{14}$     & 2s$^2$ 2p$^2$ 3d$^0$   & 1.6  & PBE & Decrease & Decrease \\
C$_{15}$     & 2s$^2$ 2p$^2$ 3d$^0$   & 1.7  & PBE & Decrease & Decrease \\
\hline \hline

\end{tabular}
\caption{The trend in the indirect and direct band gap values with 
temperature increased up to 1000 K for the different carbon (diamond) Troullier-Martins 
\textit{ab initio} pseudopotentials generated with
PBE and PZ exchange correlation functional to study finite 
temperature behavior.}
\label{ppc-g}
\end{table}

From Table~\ref{ppc-g}, the pseudopotentials C$_1$ – C$_4$ and 
C$_8$ – C$_{11}$ generated with a r$_c$ of 1.5. a.u. showed the expected 
trend of decreasing direct and indirect 
band gaps with increasing temperature. However, the band gap shifts in the two 
sets, C$_1$ – C$_4$ and C$_8$ – C$_{11}$, differ by about 4-7 meV. 
We note that the pseudopotentials C$_5$ – C$_7$, that differ from C$_1$ – C$_3$ only 
in the cut-off radius, are not temperature transferable highlighting the sensitivity to this parameter. 

\begin{figure}
\includegraphics[scale=0.46]{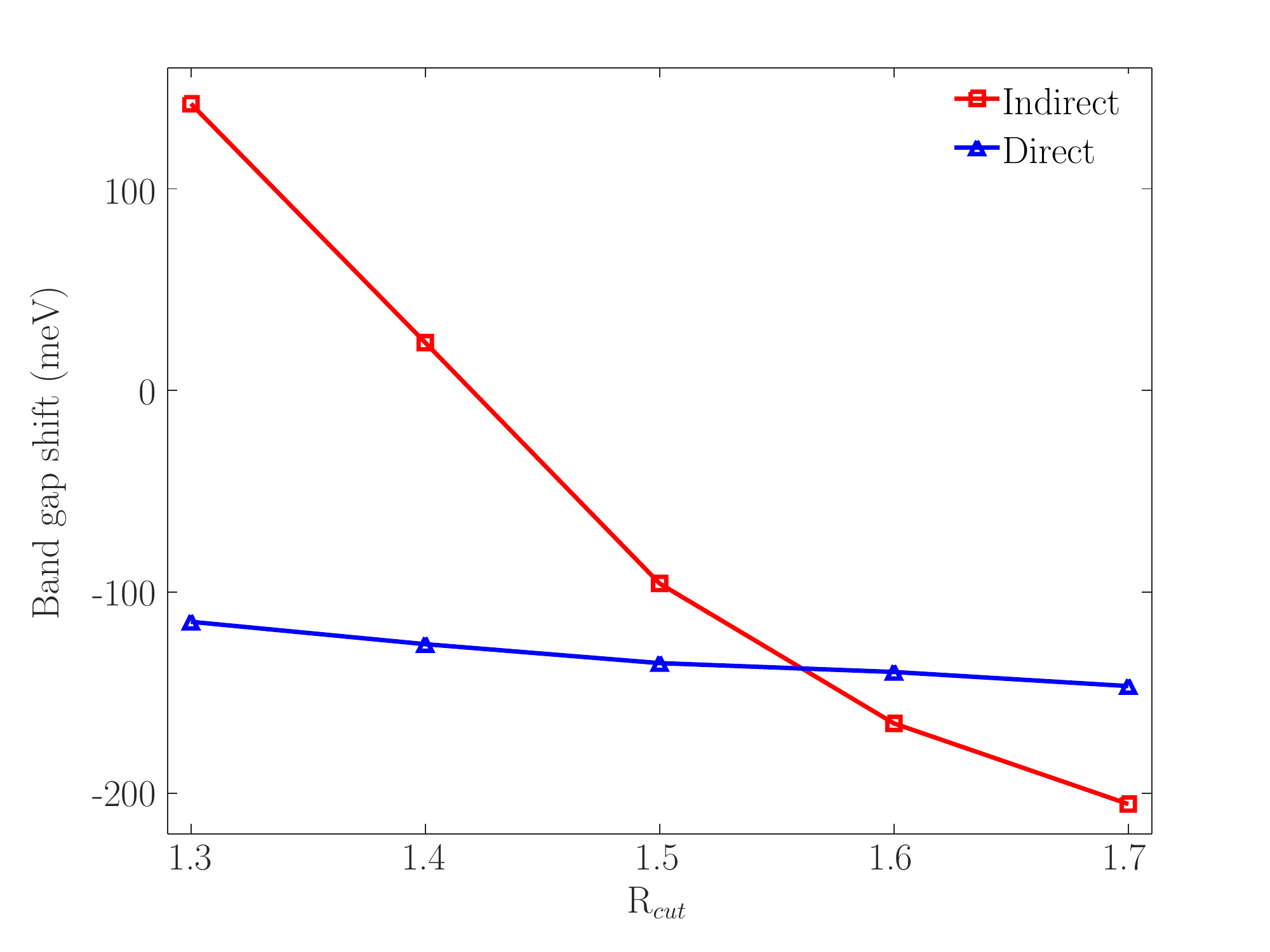}
\caption{Variation of zero-point band-gap shifts with cut-off radius for diamond } \label{CarbonRcutPlot}
\end{figure}

To further examine the sensitivity to the cut-off radius, pseudopotentials, 
C$_{12}$ - C$_{15}$ were generated with cut-off radius varying from 1.3 to 
1.7 a.u. From Table~\ref{ppc-g} it is seen that the correct trends are seen 
only for cut-off radius of 1.5 a.u. and above. Figure~\ref{CarbonRcutPlot} 
plots the variation of the zero-point indirect and direct band gap shifts 
as a function of the cut-off radius for the carbon pseudopotentials, 
generated in the 2s$^2$ 2p$^2$ 3d$^0$ configuration with the PBE 
exchange-correlation functional.  

The zero-point direct band gap shift decreases weakly with the cut-off radius. 
This is an important result for comparison of the Allen and the Allen-Heine
theories as discussed in Sec. \ref{zp-bgs}. 
In contrast, the indirect zero-point band gap
shift varies strongly with the cut-off radius. The zero-point band gap shift is 
positive for r$_c$ $<$ 1.5 a.u. leading to an increase in the indirect band gap.
Hence, carbon pseudopotentials with r$_c$ $<$ 1.5 a.u. do not exhibit 
temperature transferability. Only carbon pseudopotentials with r$_c$ $>$ 1.5 a.u. exhibit
temperature transferability. 

For silicon, from Table~\ref{ppsi-g}, pseudopotentials Si$_8$ – Si$_{10}$, 
Si$_{pbe-gipaw}$ and Si$_{12}$ - Si$_{16}$ give the 
correct band gap trend with temperature. 
The pseudopotentials Si$_1$-Si$_4$ have the same cut-off radius (1.7 a.u.) but 
different valence states and their occupancies. These pseudopotentials give 
only the zero-point indirect band gap to be larger than that of the static 
lattice. 
For all other temperatures, the indirect band gaps decrease with increasing 
temperature and are smaller than that of the the static lattice. 
In the case of the direct band gap, the zero-point band gaps are larger than 
that of the static lattice. For all other temperatures, the direct 
band gaps decrease in comparison with the zero-point vibration gap but 
are higher than that of the the static lattice up to temperature 
of 800 K.

\begin{table}[!h]
\begin{tabular}{llccll} \hline
PP & \multicolumn{3}{c} {Parameters} & \multicolumn{2}{c}{Band gap} \\ \cline{2-4}
Name &  ~Valence & ~r$_c$ & ~~~E$_{xc}$ & \multicolumn{2}{c}{Finite T trend}\\ \cline{5-6}
     &  ~electrons   & ~(a.u.)  &   & Indirect & Direct \\ \hline 
Si$_1$  & 3s$^2$ 3p$^2$            & 1.7 & PBE &  Inc-Dec  & Inc-Dec \\
Si$_2$  & 3s$^2$ 3p$^{1.5}$        & 1.7 & PBE &  Inc-Dec  & Inc-Dec \\
Si$_3$  & 3s$^2$ 3p$^1$            & 1.7 & PBE &  Inc-Dec  & Inc-Dec \\
Si$_4$  & 3s$^2$ 3p$^2$ 3d$^0$     & 1.7 & PBE &  Inc-Dec  & Inc-Dec \\
Si$_5$  & 3s$^2$ 3p$^2$            & 2.0 & PBE &  Decrease & Inc-Dec \\
Si$_6$  & 3s$^2$ 3p$^{1.5}$        & 2.0 & PBE &  Decrease & Inc-Dec \\
Si$_7$  & 3s$^2$ 3p$^1$            & 2.0 & PBE &  Decrease & Inc-Dec \\
Si$_8$  & 3s$^2$ 3p$^2$ 3d$^0$     & 2.0 & PBE &  Decrease & Decrease \\
Si$_9$  & 3s$^2$ 3p$^{1.5}$ 3d$^0$ & 2.0 & PBE &  Decrease & Decrease \\
Si$_{10}$ & 3s$^2$ 3p$^1$ 3d$^0$   & 2.0 & PBE &  Decrease & Decrease \\
Si$_{pbe-gipaw}$ & 3s$^2$ 3p$^1$ 3d$^0$   & 2.2 & PBE & Decrease & Decrease \\
Si$_{12}$ & 3s$^2$ 3p$^1$ 3d$^0$   & 2.2 & PBE &  Decrease & Decrease \\
Si$_{13}$ & 3s$^2$ 3p$^2$ 3d$^0$   & 1.8 & PBE &  Decrease & Decrease \\
Si$_{14}$ & 3s$^2$ 3p$^2$ 3d$^0$   & 1.9 & PBE &  Decrease & Decrease \\
Si$_{15}$ & 3s$^2$ 3p$^2$ 3d$^0$   & 2.1 & PBE &  Decrease & Decrease \\
Si$_{16}$ & 3s$^2$ 3p$^2$ 3d$^0$   & 2.2 & PBE &  Decrease & Decrease \\
         \hline \hline

\end{tabular}
\caption{The trend in the indirect and direct band gap values with 
temperature increased up to 1000 K for the different silicon Troullier-Martins 
\textit{ab initio} pseudopotentials generated with PBE 
exchange correlation functional to study finite temperature 
behavior.}
\label{ppsi-g}
\end{table}

For the pseudopotentials Si$_5$-Si$_7$, that were generated with a  
cutoff radius of 2.0 a.u, the zero-point indirect band gap shows the correct 
behaviour i.e. smaller than the indirect band gap of the static lattice. 
The trend in the direct band gaps however is the same as that of 
Si$_1$-Si$_4$ pseudopotentials up to a temperature of 600 K. Thus, the
pseudopotentials Si$_1$-Si$_7$ do not exhibit temperature transferability.

\begin{figure}
\includegraphics[scale=0.46]{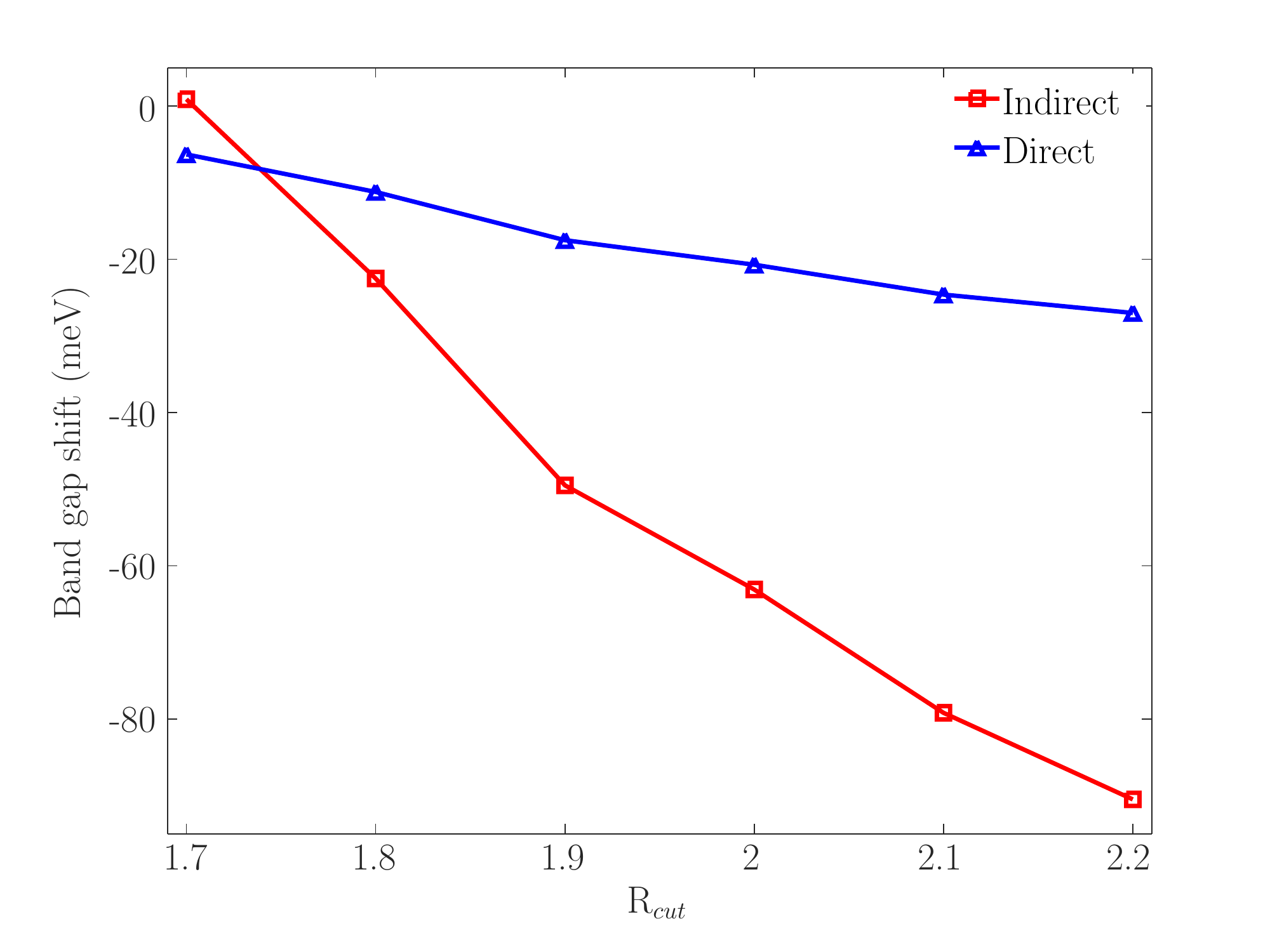}
\caption{Variation of zero-point band-gap shifts with cut-off radius for silicon} \label{SiliconRcutPlot}
\end{figure}

Table~\ref{ppsi-g} shows that the pseudopotentials Si$_8$-Si$_{10}$ differ from 
pseudopotentials Si$_5$-Si$_7$ in that the unbound 3d$^0$ state was 
incorporated in the generation condition. These pseudopotentials show the 
correct indirect and direct band gap behavior and are hence, temperature 
transferable. This leads to the conclusion that incorporation of the unbound 
3d$^0$ state in the pseudopotential generation configuration is essential 
for temperature transferability of Si pseudopotentials. 
In addition, the pseudopotential Si$_{12}$ was generated with the same 
conditions as Si$_{pbe-gipaw}$ except that it did not 
have the gipaw construction. Both Si$_{12}$ and Si$_{pbe-gipaw}$ (Table ~\ref{si-gipaw-bg})
 give virtually identical indirect and direct band gap 
shifts at all temperatures. This suggests that the gipaw construction does 
not affect temperature transferability.

To examine the dependence of the band gap shifts on the cut-off radius, 
pseudopotentials Si$_{13}$ - Si$_{16}$ were generated in the
3s$^2$ 3p$^2$ 3d$^0$  configuration with the cut-off radius 
varying from 1.8 a.u to 2.2 a.u. Figure~\ref{SiliconRcutPlot} plots the 
zero-point indirect and direct band gap shifts as a function of the cut-off 
radius.
It is again seen that the zero-point indirect band gap shift varies strongly 
with the cut-off radius. In comparison, the zero-point direct band gap shift 
varies much more slowly with the cut-off radius.

To summarize, the temperature transferability of Troullier-Martins 
pseudopotentials is dependent on the choice of the cut-off radius and the 
inclusion/exclusion of the unbound 3d$^0$ 
state in the generation configuration. The cut-off radius strongly affects 
the indirect band gap.

These observations in combination with the usual criteria for pseudopotential 
generation indicate that only in a small subset of the generation parameter 
space of Troullier-Martins pseudopotentials the new criterion of temperature 
transferability is also satisfied. These observations will also be helpful in 
generating temperature transferable pseudopotentials for other elements that 
are constituents of compound semiconductors.

The question of whether other category of pseudopotentials, Optimized, 
Ultrasoft and ONCV pseudopotentials can also be generated to exhibit
temperature transferability needs to be explored.

\subsection{Finite temperature band structure - Diamond}\lbl{tbs-c}

We calculated the finite temperature electronic structure of diamond (a 
wide band gap material) at various temperatures listed in Table~\ref{msdv} 
from 0 – 1000 K by using the DW corrected pseudopotential form factors. 
While the calculations were performed for all the 
pseudopotentials listed in Table~\ref{ppc-g}, we discuss here the results 
obtained for pseudopotentials that exhibit the correct band gap trends.

\begin{table}[!h]
\begin{tabular}{lcllrc} \hline \hline
%\begin{tabular}{l|c|ll|rc} \hline \hline
PP & a  & \multicolumn{4}{c} {Band gap} \\ \cline{3-6}
Name                &        & \multicolumn{2}{c}{Static lattice} & \multicolumn{2}{c}{Zero point shift} \\ 
                    & (a.u.) & \multicolumn{2}{c}{(eV)} & \multicolumn{2}{c}{(meV)} \\
            &        &  Indirect  &  Direct  & Indirect & Direct  \\ \hline 
C$_1$    & 6.74 & ~4.26  & 5.634 & ~102 & 140 \\
C$_2$    & 6.73 & ~4.28  & 5.635 & ~102 & 142 \\
C$_3$    & 6.71 & ~4.33  & 5.66  & ~103 & 147 \\
C$_4$    & 6.66 & ~4.31  & 5.68  & ~107 & 149 \\
C$_8$    & 6.72 & ~4.215 & 5.62  & ~96  & 135 \\
C$_9$    & 6.71 & ~4.247 & 5.64  & ~96  & 138 \\
C$_{10}$ & 6.70 & ~4.29  & 5.654 & ~95  & 141 \\
C$_{11}$ & 6.64 & ~4.27  & 5.68  & ~100 & 143 \\
C$_{14}$ & 6.70 & ~4.25  & 5.65  & ~165 & 140 \\
C$_{15}$ & 6.68 & ~4.29  & 5.67  & ~205 & 146 \\
\hline \hline

\end{tabular}
\caption{The lattice constant, indirect and direct band gaps for the static
lattice and zero-point energy shifts in the direct and indirect band
gaps for diamond obtained from temperature transferable pseudopotentials}
\label{c-bg-0}
\end{table}

Table~\ref{c-bg-0} gives the zero-point shifts for the pseudopotentials 
that exhibited temperature transferability. 
It shows that the zero-point band gap shifts fall into two 
clear sets for C$_1$-C$_4$ and C$_8$-C$_{11}$ where the former gives slightly 
higher shifts of 4-7 meV. All these pseudopotentials have the same cut-off 
radius (1.5 a.u).

The main difference between the two sets is the 
incorporation of the unbound 3d$^0$ state in the pseudopotential generation 
configuration in C$_8$-C$_{11}$. Thus, for the case of diamond, the 
incorporation of the unbound 3d$^0$ state in the generation configuration 
has a consistent, but minor, effect.

The pseudopotentials C$_{14}$ and C$_{15}$ vary from C$_8$ only in the cut-off 
radius. The choice of the cut-off radius strongly influences the indirect 
zero-point band gap shifts. The direct zero-point band gap shifts vary weakly 
with the cut-off radius. Comparing with the previous results~\cite{GiusCohen2010, PonceCMSc2014}, based on the 
Allen-Heine theory, for the direct and indirect band gap shifts due to the DW 
term, the pseudopotentials we generated with the cut-off radius of 1.5 a.u. 
give the closest agreement. Hence, we restrict our discussion to the results 
obtained with C$_8$ and C$_{11}$ pseudopotentials that differ in the 
generation parameters only in the exchange-correlation functional.

For diamond, the lattice constants and the 
static lattice band gaps are very similar to that in previous 
\textit{ab initio} studies~\cite{GiusCohen2010, PonceCMSc2014, Gonze2014, 
PatrickGiustino2014, ZachGius2016, Ponce2015, MonserratNeeds2013, 
MonserratNeeds2014}. 
The band gaps are lower than the experimental values due to the well known 
underestimation of band gaps in density functional theory~\cite{GiusCohen2010, 
PonceCMSc2014, Gonze2014, PatrickGiustino2014, ZachGius2016, Ponce2015, MonserratNeeds2013,
 MonserratNeeds2014}. 

From Table~\ref{c-bg-0} it is seen that the lattice parameters for 
pseudopotentials with the LDA(PZ) parameterization (C$_4$ and C$_{11}$) are 
lower than others with the GGA(PBE) parameterization.
In addition, the zero-point band gap shifts for pseudopotentials with 
the LDA(PZ) parameterization C$_4$ and C$_{11}$ (Table~\ref{c-bg-0}) are higher than 
others with the GGA(PBE) parameterization. Both these trends are similar to 
results from previous studies~\cite{PonceCMSc2014}.
 In particular, the lattice parameter for the pseudopotential C$_{11}$ is 
6.64 a.u. and is very similar to 3.52 \AA~\cite{GiusCohen2010} and 
6.652 Bohrs~\cite{PonceCMSc2014} for the pseudopotential with the same
 generation conditions. In addition, the direct band gap for 
 the pseudopotential C$_{11}$ is 5.68 eV that is very close to the value of 
5.67 eV for the similar `reference' pseudopotential in 
Ponce \textit{et al.}~\cite{PonceCMSc2014}. 
The similarities on the lattice parameter and band gap for the 
pseudopotential C$_{11}$ with earlier results~\cite{GiusCohen2010, 
PonceCMSc2014} justifies further comparisons.

\begin{figure}
\includegraphics[scale=0.36,angle=270]{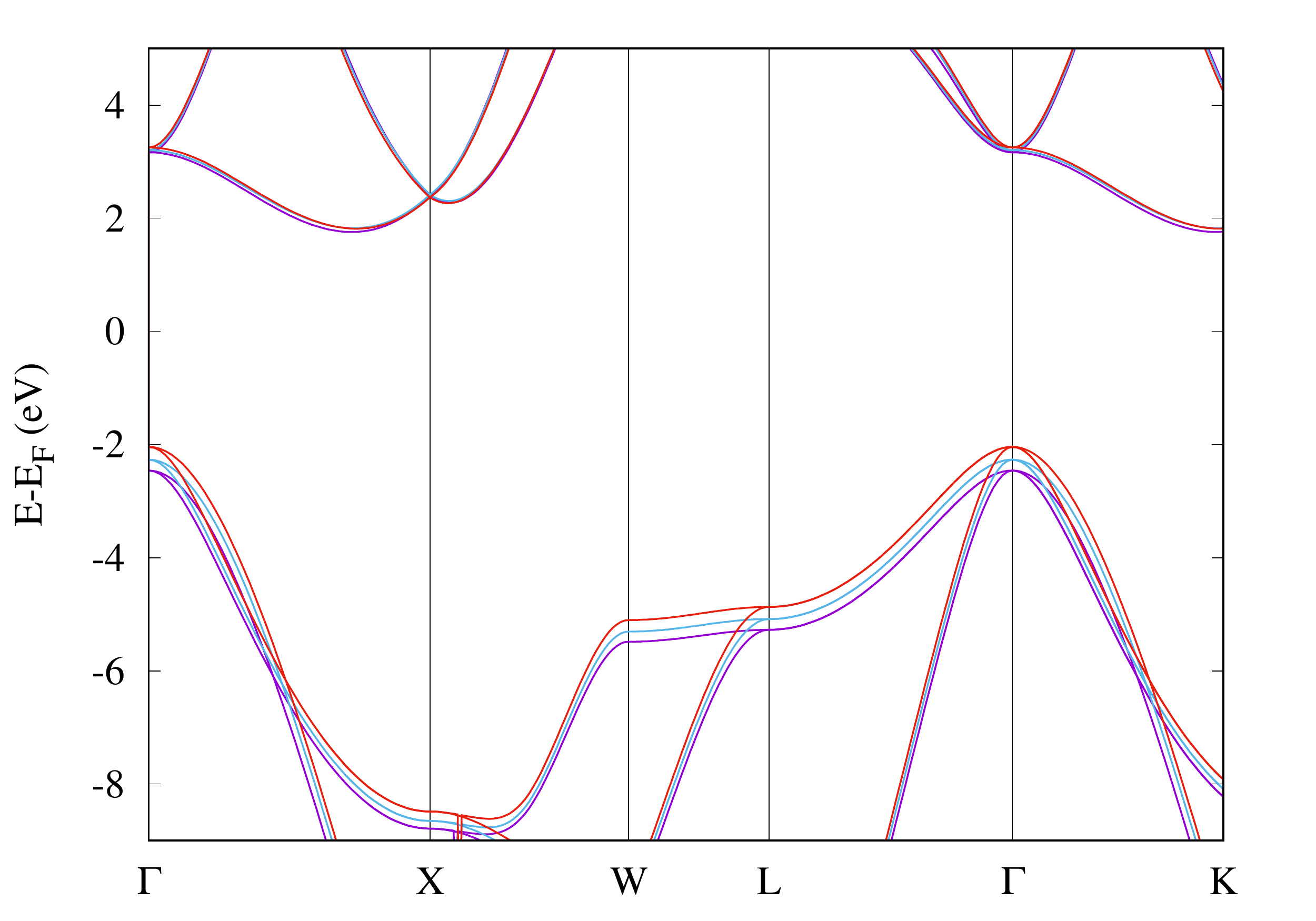}

\caption{The calculated band structure (static lattice and thermally-averaged)
 of diamond at 0 K, 300 K and 1000 K in the vicinity of the Fermi level. The violet 
 line represents the band structure at 0 K, the blue line at 300 K and the 
red line at 1000 K.}
\label{c-bs}
\end{figure}

Figure~\ref{c-bs} plots the static lattice (0 K) band structure of diamond 
and also the thermally-averaged band structures (at 300 K and 1000 K) for the 
pseudopotential C$_8$ for the in the vicinity of the Fermi level.
The band structure is plotted so that the Fermi level at each temperature
is at 0 eV. Fig.~\ref{c-bs} demonstrates that the finite temperature thermally-averaged
band structure can be obtained via  an \textit{ab initio} 
electronic structure calculation using \textit{ab initio} pseudopotentials using 
the Allen theory formalism. The finite temperature band structures show that the 
valence and conduction bands shift by different amounts for different k points. 
From Fig.~\ref{c-bs} it is clear that the temperature dependent 
band gap shifts can be obtained for any k point.

Earlier \textit{ab initio} studies~\cite{GiusCohen2010, PonceCMSc2014, Gonze2014, 
PatrickGiustino2014, ZachGius2016, Ponce2015, MonserratNeeds2013, 
MonserratNeeds2014}, based on the Allen-Heine theory, report the band 
gap shifts only at special k points, corresponding to the direct and 
indirect band gap shifts. 
The temperature dependent band structure of diamond has been obtained
with path-integral molecular dynamics simulations~\cite{Ramirez}. 
 These simulations are computationally intensive as the ensemble-averages 
are calculated over $\sim$ 10$^5$ steps~\cite{Ramirez}. Our results 
highlight the advantages of implementing the Allen theory to obtain 
finite temperature band structures without any increase in the 
computational expense.

\begin{figure}
\includegraphics[scale=0.36,angle=270]{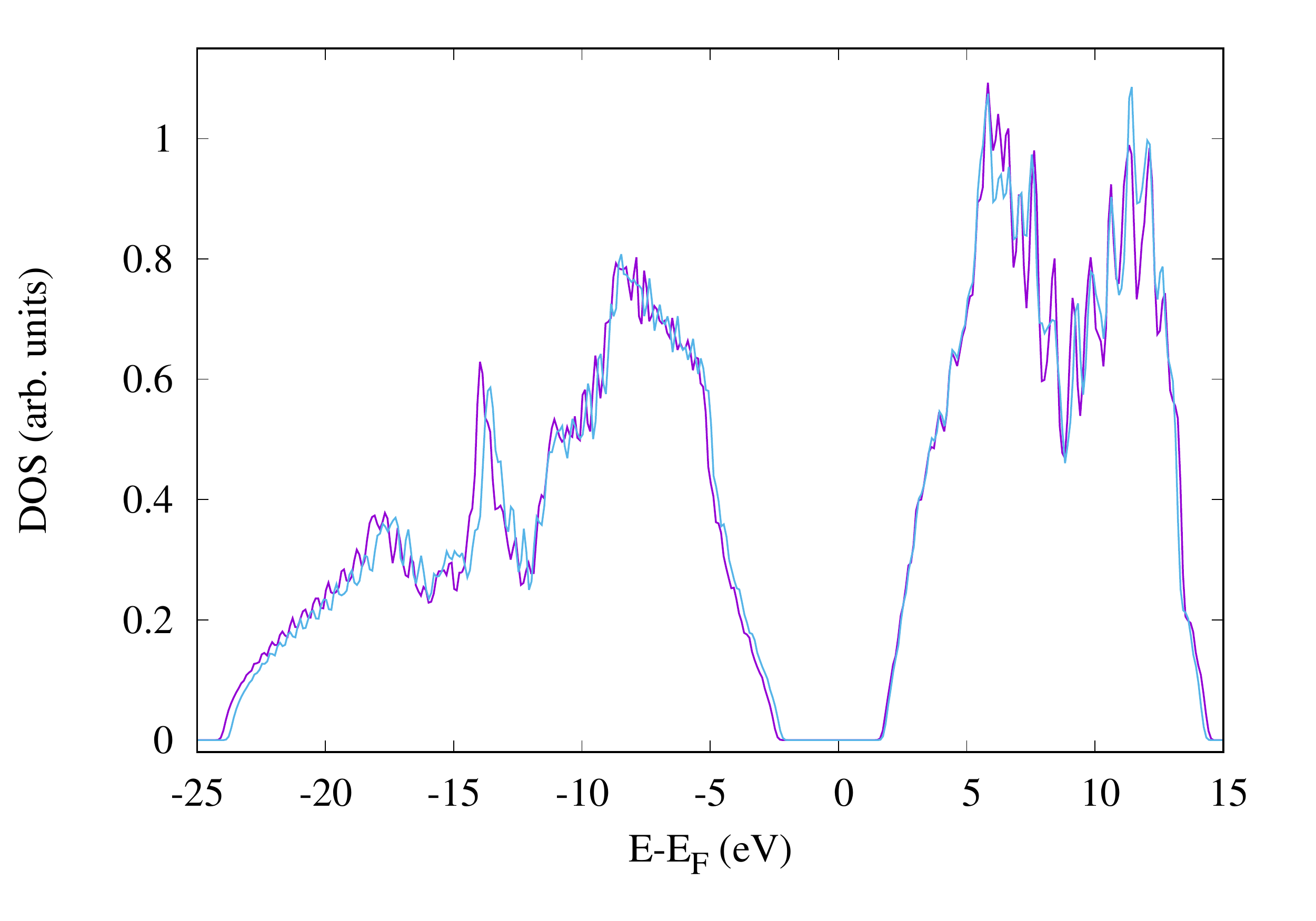}

\caption{The density of states of diamond at 0 K (violet) and 300 K (blue)}
\label{c-dos}
\end{figure}

To understand the effect of temperature on the complete band
structure we plot the density of states (DOS) for the carbon 
pseudopotential C$_8$ at 0 K and 300 K in Figure~\ref{c-dos}. 
The effect of temperature on the valence electron 
energy diagram is noticeable from the change in the DOS at 300 K. 
The DW term causes a positive shift of the valence electron energy levels 
towards the Fermi energy. The unfilled conduction band energy levels 
on the other hand experience a negative shift towards the Fermi energy. 
This feature is observed in the DOS of all finite temperature band structures 
that we have calculated. In the filled as well as unfilled electron energy 
levels, the amount of shift experienced from the static lattice is different 
for different levels indicating that the levels are not just scaled uniformly 
by a temperature dependent scaling constant. 

\begin{figure}
\includegraphics[scale=0.47]{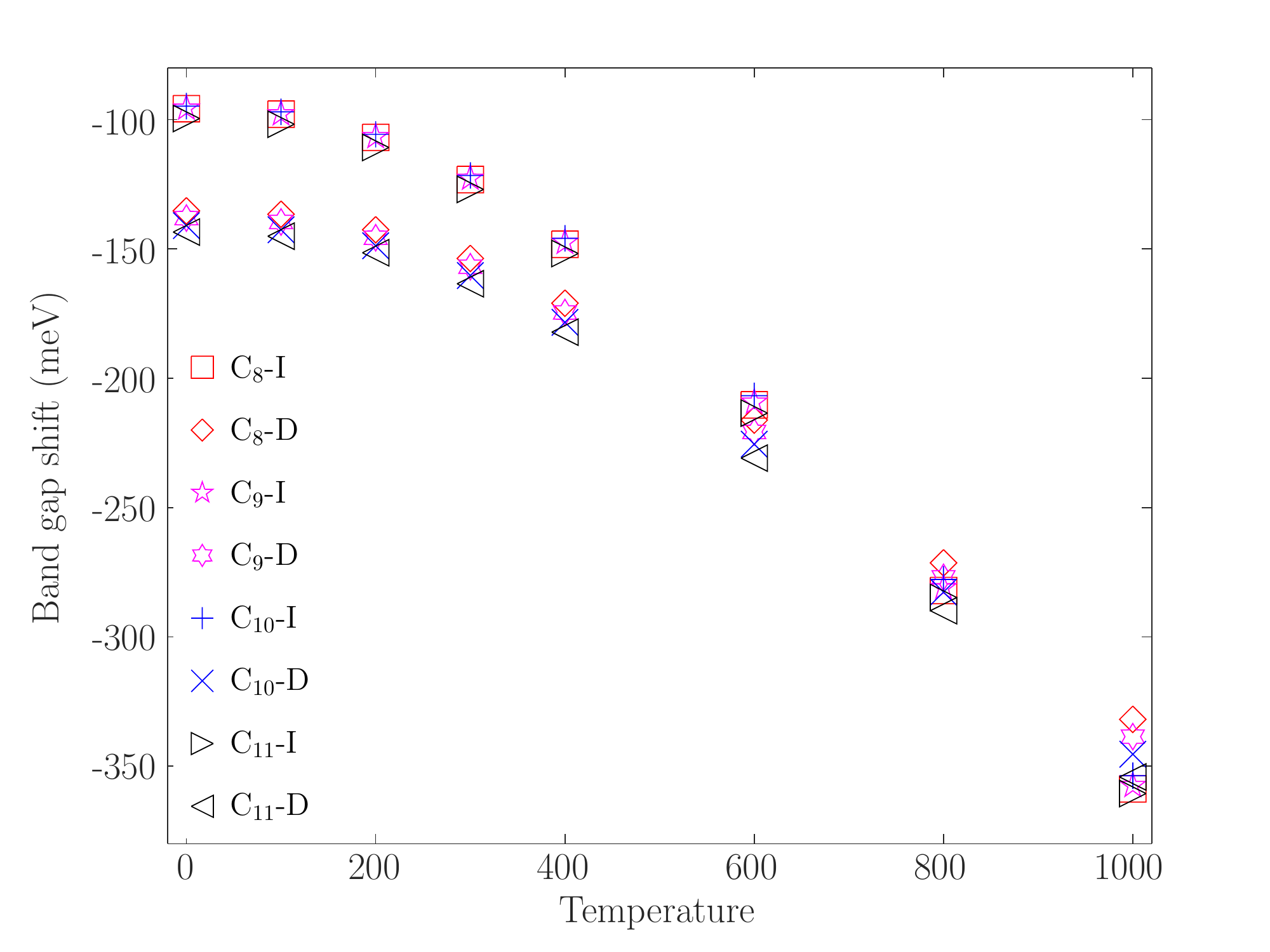}

\caption{The direct (D) and indirect (I) band gap shifts in diamond due to the DW term with
increasing temperature for different temperature transferable carbon 
pseudopotentials}
\label{c-bg-shift}
\end{figure}

Figure~\ref{c-bg-shift} plots the indirect and direct band gap shifts of diamond due to 
the DW term for the four carbon pseudopotentials, C$_8$-C$_{11}$, at various 
temperatures. The indirect and direct band gap shifts show a non-linear 
behavior at low temperatures and a linear behavior 
at higher temperatures consistent with experimental observations~\cite{Ponce2015}.
While we have calculated the shifts for all pseudopotentials
at all temperatures under investigation, we plot the results only for the 
pseudopotentials that exhibit good temperature transferability behavior and
hence are of interest for further studies. 
We note that the indirect band gap shifts decrease more rapidly with 
temperature than the direct band gap shifts and the changes in 
the generation configuration, in terms of the occupancies of the 2p state, 
has only a minor effect at all temperatures. 

\subsubsection{Zero-point band gap shifts}\lbl{zp-bgs}

Zero-point vibrations represent the minimum perturbation to the static 
lattice system. Hence, it is of interest to examine the zero point band gap 
shifts in detail.
For zero-point shifts, we assume that the DW term can be approximated by the 
second order expansion and the neglect of higher order terms in Allen-Heine 
theory is unimportant. That is, V$_G$ e$^{-W} \approx $ V$_G$ (1-W) and the 
Allen and Allen-Heine theories should give similar zero-point band gap 
shifts.

Of the several \textit{ab initio} studies~\cite{GiusCohen2010, PonceCMSc2014, Gonze2014, 
PatrickGiustino2014, ZachGius2016, Ponce2015, MonserratNeeds2013, MonserratNeeds2014} 
on the band gap shifts in diamond based on Allen-Heine theory, only 
two~\cite{GiusCohen2010, PonceCMSc2014} give separately the contributions of 
the DW and the SE terms. Hence, the DW contribution to the zero-point shifts in our study (Table~\ref{c-bg-0}) can 
be compared with only these studies~\cite{GiusCohen2010, PonceCMSc2014}. In particular, our C$_{11}$ 
pseudopotential has the same generation conditions as the pseudopotential 
in Giustino \textit{et al.}~\cite{GiusCohen2010} and the `reference' pseudopotential in 
Ponce \textit{et al.}~\cite{PonceCMSc2014}. These studies, however, report the DW contribution 
only to the direct band gap and hence, the comparison is restricted to 
it.  

The DW contribution to the direct band gap shifts for the  pseudopotentials 
in the present study range from 135-149 meV for
a variety of pseudopotential generation conditions (Table~\ref{c-bg-0}). 
In contrast, the indirect band gap shifts of these pseudopotentials (Table~\ref{c-bg-0})
 have a much wider spread (95-205 meV) on account of the sensitivity to 
cut-off radius. For the direct as well as indirect band gap shifts, the 
pseudopotentials generated with PBE exchange correlation functional and
that include the 3d$^0$ state give comparatively lower shifts for a constant
cut-off radius. When the direct band gap shifts are compared to the literature 
values~\cite{GiusCohen2010, PonceCMSc2014} of 105-121 meV for the zero-point shifts in the Allen-Heine 
theory, for Troullier-Martins pseudopotentials with a variety of generation 
conditions, our results are higher by about 20-30 meV.

In particular, Giustino \textit{et. al}~\cite{GiusCohen2010} report a 
zero-point shift of 117 meV (19\% of 615 meV) due to the DW term for the 
direct band gap of diamond.  Poncé \textit{et. al}~\cite{PonceCMSc2014} report 
that the DW contribution ranges from 105-121 meV for four (Troullier-Martins) 
pseudopotentials. Of these, they report~\cite{PonceCMSc2014} a direct band gap 
shift of 115 meV for the `reference' pseudopotential, with the same generation 
conditions as in Giustino \textit{et. al}~\cite{GiusCohen2010} and in C$_{11}$ 
in the present study. This is virtually identical to the DW zero-point shift of
 117 meV reported in Giustino \textit{et. al}~\cite{GiusCohen2010}. 
Thus, the differences in the total (DW+SE) zero-point shifts between Giustino \textit{et. al}~\cite{GiusCohen2010} 
and Poncé \textit{et. al}~\cite{PonceCMSc2014} is due to the SE contribution. 
The DW contribution to the zero-point direct band gap shift for the C$_{11}$ 
pseudopotential in the present study is 143 meV. This value can be directly 
compared and is $\sim$ 26 meV higher than that reported in the above 
studies~\cite{GiusCohen2010, PonceCMSc2014}.

\subsubsection{Finite temperature band gap shifts}\lbl{ht-bgs}

The DW contribution to the band gap shifts calculated in the present study can 
also be compared with results from Allen-Heine 
theory at high temperatures, where the higher order terms may become important. 
We compare the the DW contribution to the direct band gap shifts obtained with the C$_{11}$ 
pseudopotential since it has the same generation configuration
as in literature~\cite{GiusCohen2010}.

Our calculated DW band gap shift of 152 meV at 200 K, 164 meV at 300 K
and 231 meV at 600 K compares well with the reported DW contribution 
to the direct band gap shift of $\sim$ 160 meV at 200 K, 190 meV at 300 K and
 280 meV at 600K in literature~\cite{GiusCohen2010}. However, an inversion must be noted. 
 Our finite temperature DW band gap shifts are smaller while our zero-point shifts are larger 
 compared to literature values~\cite{GiusCohen2010, PonceCMSc2014}.

As seen in the previous section, there is a difference of $\sim$ 26 meV between the 
results obtained using the Allen and Allen-Heine theory in the DW 
zero-point band gap shifts. Further insight can be gained if the differences 
are compared with respect to the zero-point shifts which eliminates the effect 
of the different zero-point band gap shifts.  This helps to examine the 
increase in the band gap shifts due to the increase in the mean-square 
displacements with temperature.
 
Eliminating the zero-point shifts, the additional DW band gap shift in 
our study is 9 meV at 200 K, 21 meV at 300 K and 87 meV at 
600 K. These small increases in the band gap shifts with temperature can be 
attributed to the fact that the $\langle u_i^2 \rangle $ changes very slowly 
with temperature for diamond (Table~\ref{msdv}). For example, $\langle u_i^2 \rangle $ 
varies from 0.00161 {\AA$^{2}$} (0.001 K) to 0.00169 {\AA$^{2}$} (200 K) 
to 0.0018 {\AA$^{2}$} (300 K) at low temperatures. 
 
In comparison, in Giustino \textit{et. al}~\cite{GiusCohen2010}, after eliminating the zero-point shifts, the 
additional DW band gap shifts are $\sim$ 40 meV at 200 K, 70 meV at 300 K and 
160 meV at 600 K. Compared to our results above, the 
relatively large band gap shifts for very small increases in the mean-square displacement
 values (especially upto 300 K) appear excessive. 

\subsection{Silicon}\lbl{si}

Silicon, a small band gap semiconductor used  extensively in a wide
range of technology applications is well studied experimentally as 
well as theoretically and is of interest for its temperature dependent properties.
We calculated the finite temperature thermally-averaged electronic structure of 
silicon at various temperatures listed in Table~\ref{msdv} by 
using the DW factor modified pseudopotential. 
We performed the finite temperature calculations for all pseudopotentials
listed in Table~\ref{ppsi-g}.  

Table~\ref{si-bg} lists the zero-point shifts in the indirect and direct
band gap for all the temperature transferable pseudopotentials. 
It shows that pseudopotentials Si$_8$-Si$_{10}$, which differ 
only in the valence state 
occupancies of p electrons, have similar zero-point shifts.
The band gap shifts are virtually identical for Si$_{pbe-gipaw}$ 
(Table ~\ref{si-gipaw-bg}) and Si$_{12}$ not only for zero-point vibrations 
but also for all other temperatures in our study. 
The band gap shifts increase monotonically for Si$_{13}$ - Si$_{16}$ indicating a 
strong dependence on the cut-off radius as seen 
in Figure~\ref{SiliconRcutPlot}.

\begin{table}[!h]
\begin{tabular}{lcllll} \hline \hline
PP & a  & \multicolumn{4}{c} {Band gap} \\ \cline{3-6}
Name                &        & \multicolumn{2}{c}{Static lattice} & \multicolumn{2}{c}{Zero point shift} \\
                    & (a.u.) & \multicolumn{2}{c}{(eV)} & \multicolumn{2}{c}{(meV)} \\
            &        &  Indirect  &  Direct  & Indirect & Direct  \\ \hline
Si$_8$     & 10.33  & ~0.631  & 2.565  & ~63.1  & 20.7 \\
Si$_9$     & 10.33  & ~0.633  & 2.567  & ~64.3  & 20.8 \\
Si$_{10}$  & 10.33  & ~0.636  & 2.568  & ~65.3  & 21 \\
Si$_{pbe-gipaw}$ & 10.31  & ~0.633  & 2.572  & ~94    & 28 \\
Si$_{12}$  & 10.31  & ~0.633  & 2.572  & ~93  & 27 \\
Si$_{13}$  & 10.34  & ~0.633  & 2.565  & ~23  & 11 \\
Si$_{14}$  & 10.33  & ~0.633  & 2.565  & ~50  & 18 \\
Si$_{15}$  & 10.32  & ~0.627  & 2.566  & ~79  & 24.6 \\
Si$_{16}$  & 10.31  & ~0.624  & 2.566  & ~91  & 27 \\
\hline \hline
\end{tabular}
\caption{The lattice constant, indirect and direct band gaps for the static
lattice and zero-point direct and indirect band gap shifts
for silicon obtained from temperature transferable pseudopotentials}
\label{si-bg}
\end{table}

The static lattice direct (2.56 eV) and indirect  (0.63 eV)
band gaps for these pseudopotentials are very close to literature 
values~\cite{PatrickGiustino2014} of 2.55 eV and 0.62 eV respectively. 
Si$_{12}$, Si$_{pbe-gipaw}$, Si$_{16}$ with higher 
cut-off radius have higher indirect band gap shifts and 
Si$_{13}$, Si$_{14}$ with a lower cut-off radius have lower band gap 
shifts. As observed in diamond, the indirect band gap shifts are 
highly sensitive to the cut-off radius used in the pseudopotentials.

Previous literature values for the total (DW + SE) shift of the indirect 
band gap due to zero-point vibrations are 57 meV~\cite{PatrickGiustino2014, ZachGius2016}, 60 meV~\cite{MonserratNeeds2014} and
 64.3 meV~\cite{Ponce2015} and the experimental values are 62-64 meV~\cite{ZachGius2016}. For the direct band gap, the reported values are 22 meV~\cite{PatrickGiustino2014} and 47 meV~\cite{Ponce2015} and the experimental values are 25$\pm$17~\cite{ZachGius2016}. 

Previous studies~\cite{AllenCardona, Zollner}, though based on empirical 
pseudopotentials, suggest that in silicon the DW term overestimates the band 
gap shifts. The SE term is of opposite sign so that the total band gap 
shifts are less than those obtained from just the DW term.
 
The temperature transferable pseudopotentials in Table ~\ref{si-bg} 
give different magnitudes of zero-point band gap shifts than in 
literature~\cite{AllenCardona, Zollner, PatrickGiustino2014, ZachGius2016, 
MonserratNeeds2014, Ponce2015}. This is due to the strong sensitivity of the 
band gap shifts to the cut-off radius.
This comparison shows that only the pseudopotentials
 Si$_8$-Si$_{10}$ and Si$_{15}$ give, band gap shifts that are the 
closest to previous results. 

\begin{figure}
\begin{center}
\includegraphics[scale=0.47]{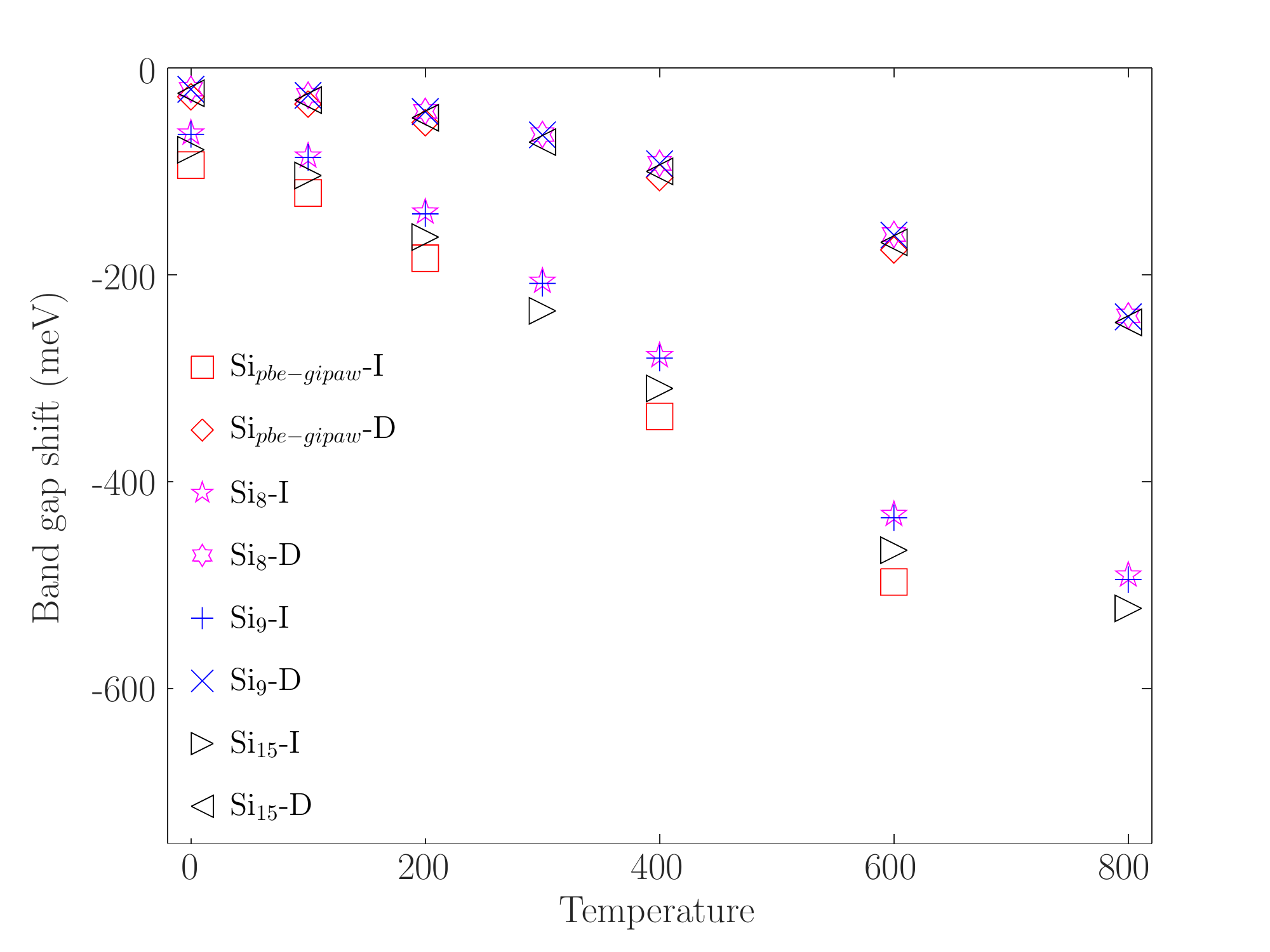}
\end{center}

\caption{The direct (D) and indirect (I) band gap shifts in silicon due to the DW term with 
increasing temperature for different temperature transferable pseudopotentials}
\label{si-bg-shift}
\end{figure}

None of the \textit{ab initio} studies on silicon separately report the DW contribution. 
However, Allen and Cardona~\cite{AllenCardona}, in their study based on the Allen-Heine 
theory with empirical pseudopotentials, report the DW contribution to 
the direct band gap shift at different temperatures upto 600 K. 
Our DW direct band gaps shifts (Figure~\ref{si-bg-shift}) in the
temperature range upto 600 K are very similar to these reported values~\cite{AllenCardona}.
Allen and Cardona~\cite{AllenCardona}, however, report negligible zero-point direct 
band gap shift, while we report a finite value.  

Figure~\ref{si-bg-shift} plots the indirect and direct band gap shifts for 
the four silicon pseudopotentials, Si$_8$, Si$_9$, Si$_{pbe-gipaw}$ and 
Si$_{15}$, at various temperatures. The indirect and direct band gap shifts 
exhibit a nonlinear behavior at low temperatures and a linear behavior 
at higher temperatures as seen in experimental studies~\cite{Ponce2015}.

\begin{figure}
\begin{center}
\includegraphics[scale=0.36,angle=270]{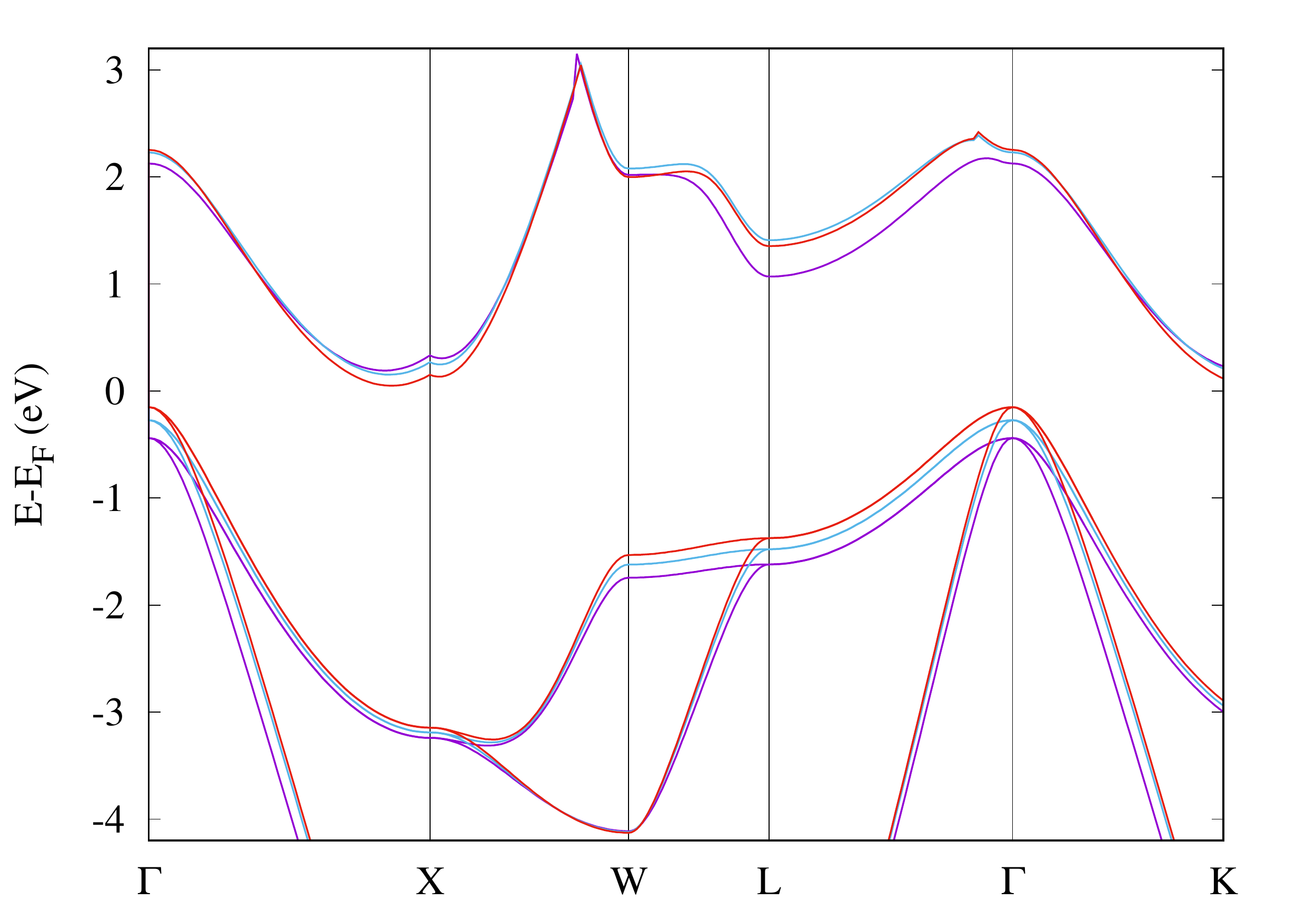}
\end{center}
\caption{The calculated band structure (static lattice and thermally-averaged) 
of silicon at 0K, 300K and 600K in the vicinity of the Fermi level. The violet 
line represents the band structure at 0 K, the blue line at 300K and the red line at 600 K.}
\label{si-bs}
\end{figure}

The effect of temperature on Si electronic structure is investigated 
 and in Figure~\ref{si-bs} we plot the (static lattice and thermally-averaged) 
 band structure of silicon for the pseudopotential Si$_8$ for 0 K, 300 K and 600 K. 
 The effect of temperature on the Si electronic 
 structure is quite interesting with differing positive and negative shifts in the 
energy with respect to the static lattice. The energy shifts at the 
different high symmetry points in the Brillouin zone are significantly 
different for different temperatures.  The shifts in the energy values 
at different points within any high symmetry line of the Brillouin zone 
with respect to temperature are observed to vary.

\begin{figure}
\begin{center}
\includegraphics[scale=0.36,angle=270]{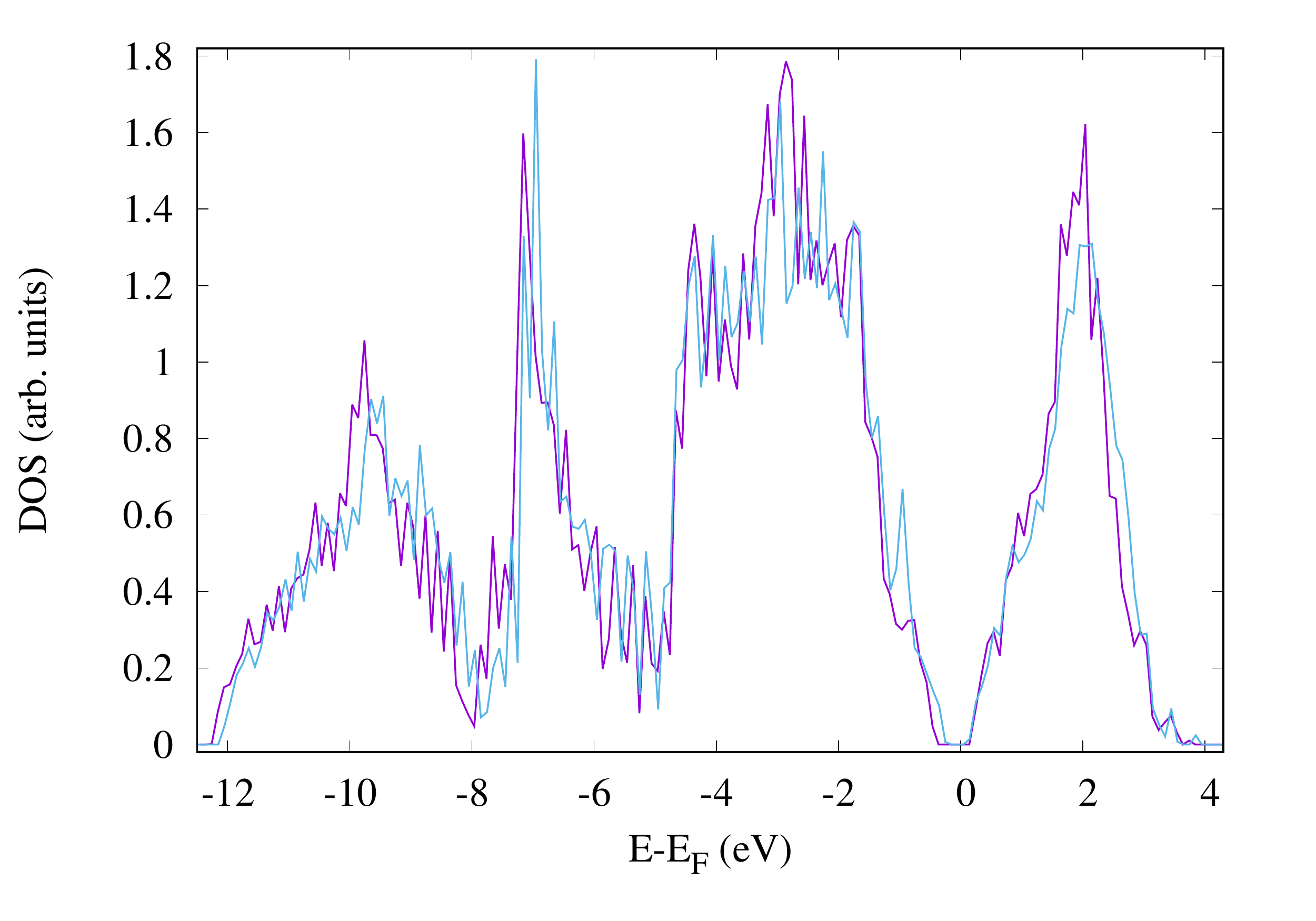}
\end{center}

\caption{The density of states of silicon at 0K (violet)  and 300K (green)}
\label{si-dos}
\end{figure}

Figure~\ref{si-dos} plots the density of states for the silicon 
pseudopotential Si$_8$ at 0 K and 300 K. As in the case of diamond, the 
temperature dependent pseudopotential affects not only the valence band 
and conduction band energies near the Fermi level but throughout the 
valence and conduction bands.
For finite temperatures, the filled bands and the unfilled bands shift 
towards the Fermi energy nonuniformly as seen in our studies on diamond. 
In general, the density of states broaden and shift towards the Fermi 
level. The amount of broadening and shift increases with increased temperature.

\subsection{Discussion of Allen and Allen-Heine theory results}\lbl{comparison}

The DW term of the electron-phonon interaction is more accurately evaluated~\cite{AllenHeine} 
in the implementation of the  Allen theory, whereas, it is approximated to the second order 
in the Allen-Heine theory. Considering this some discrepancies in the band gap shift values 
obtained from these theories is expected. However, if the second-order approximation is considered to be valid,
then both theories should give similar band gap shifts.

In order to separate the effect of the approximation to the DW term from other input parameters used in 
the calculations, it is essential that the same \textit{ab initio} 
pseudopotential and mean square displacements of the atoms be used. 
In the present study, we have used the same pseudopotential for diamond
as in previous studies~\cite{GiusCohen2010, PonceCMSc2014} based on the Allen-Heine theory. 
The mean square displacement values as in previous studies however, could 
not be used since they have not been reported.

The implementation of Allen theory is based on using explicit values of the 
mean-square displacements, whereas in the Allen-Heine theory the mean-square
displacement values are not directly used and are implicit.
In our study of finite temperature properties of diamond and silicon
explicit values of the mean-square displacements obtained from 
\textit{ab initio} studies have been used. 
If the values of the mean-square displacements are different from that
in the earlier studies, then discrepancies can be expected in the results
from the two theories.

In the case of diamond, we have used (Table~\ref{msdv}) the 
zero-point mean-square displacement values reported in Schowalter 
\textit{et al.}~\cite{Schowalter}. A similar value, for the zero-point vibrations, 
has also been reported in Yang \textit{et al.}~\cite{Yang}.  
In both these studies~\cite{Schowalter, Yang}, the mean-square displacement 
values have been obtained from density-functional perturbation theory. 
The previous Allen-Heine based studies~\cite{GiusCohen2010, PonceCMSc2014} that reported the 
DW contribution to the direct band gap shift also use the density-functional 
perturbation theory to account for thermal vibrations, though explicit values 
of the zero-point mean-square displacements are not reported. Since there are
some discrepancies in the DW band gap shifts for diamond (as seen earlier), 
it would be of interest to compare the zero-point mean-square 
displacements implicit in the Allen-Heine theory based 
studies~\cite{GiusCohen2010, PonceCMSc2014} with the explicit values 
reported in literature~\cite{Schowalter, Yang} that are used in our study
based on the Allen theory.

If the mean-square displacement values used in the two theories are the 
same, then the discrepancies, if any, are likely due to the different 
numerical implementations of the electron-phonon contribution to the 
electronic energies in the two theories.

Clearly, further studies are needed to fully understand  the results 
obtained using the Allen and Allen-Heine theories. Such studies for 
any semiconductor, however must necessarily use temperature 
transferable pseudopotentials.

\subsection{Finite temperature valence charge density}\lbl{ft-vcd}

Finite temperature charge densities are of interest, both theoretically and 
experimentally, to study bonding in materials.  At present, there is no 
simple \textit{ab initio} method to obtain finite temperature valence 
electron wavefunctions or charge densities. Currently, finite temperature 
charge densities are obtained from static lattice calculations with 
empirical approaches to incorporate the effect of thermal vibrations as 
discussed below.

The valence electron densities have been experimentally obtained for several 
materials from x-ray diffraction and electron diffraction studies~\cite{Bindzus, Wahlberg, Nakashima,Sang, Zuo, Dudarev}. 
In particular, silicon and diamond are considered to be the prototype 
materials in experimental (valence) electron density studies~\cite{Bindzus, Wahlberg}. 
In these studies, the `fundamental step'~\cite{Wahlberg} is to correct for the 
observed intensities at finite temperatures with the DW factor~\cite{Bindzus, Wahlberg, Nakashima,Sang, Zuo, Dudarev}. 
Following this correction, an effective static lattice charge density is 
obtained. Subsequently, these charge densities are compared with 
theoretical \textit{ab initio} calculations performed for 
static lattice~\cite{Bindzus, Wahlberg, Nakashima,Sang}.

\begin{figure*}[ht]
\begin{center}
{\bf (a)}

\includegraphics[scale=0.5]{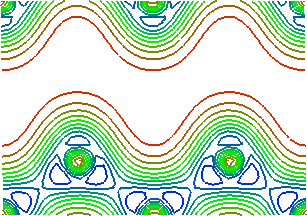} \hskip 0.1in \includegraphics[scale=0.5]{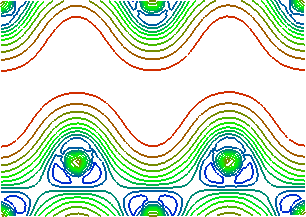} \hskip 0.1in \includegraphics[scale=0.5]{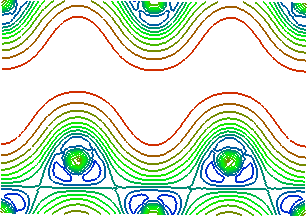}
 
\vskip 0.2in
{\bf (b)}

\includegraphics[scale=0.5]{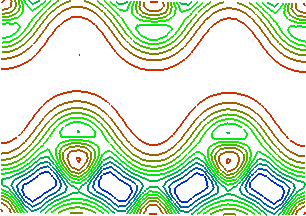} \hskip 0.1in \includegraphics[scale=0.5]{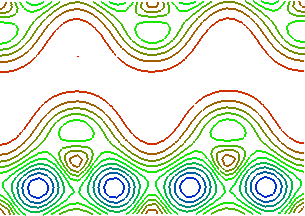} \hskip 0.1in \includegraphics[scale=0.5]{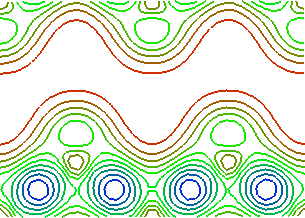}

{\bf 0 K}  \hskip 2.0in {\bf 300 K}  \hskip 2.0in  {\bf 600 K}
\end{center}

\caption{The charge density along the (110) plane for (a) diamond and 
(b) silicon.
The charge density of the static lattice (0 K) is shown in the left panel,
at 300 K in the middle panel and at 600 K in the right panel for both the
materials}

\lbl{c-si-cden}
\end{figure*}

Alternately, the theoretical charge density from static lattice calculations 
are altered by incorporating thermal vibrations through the DW factors and the finite 
temperature charge densities or structure factors so obtained are compared 
with finite temperature experimental values~\cite{Zuo, Dudarev}. 

Fundamental to these approaches is the assumption of the rigid pseudo-atom 
approximation~\cite{Gatti, Jones}, viz. the electron density, including the valence 
electron density, can be partitioned to individual atoms and each such segment 
is frozen and moves rigidly with the thermal vibrations. This leads to the 
DW factor correction, the `fundamental step' in the charge density studies. 
We note that Jones and March~\cite{Jones} had raised  concerns about the validity 
of the rigid pseudo-atom approximation for valence electrons. 

In the rigid pseudo-atom approximation, since the valence charge densities move rigidly with thermal motion, 
it implies that there is no change in the valence electron wavefunctions and energies. 
It follows that the band gaps must be unchanged with temperature. 
This is contrary to experimental and theoretical observations~\cite{Giustino2017, PoncePRB2014, AllenCardona, Zollner, Marini, GiusCohen2010, PonceCMSc2014, Gonze2014, PatrickGiustino2014, ZachGius2016, Ponce2015, MonserratNeeds2013, MonserratNeeds2014}. 
This is a simple demonstration of the incorrectness of the rigid pseudo-atom 
approximation for valence electrons in silicon and diamond. Hence, at present, 
there is no simple and theoretically justified method to obtain finite 
temperature valence charge densities. 

In the Allen theory~\cite{Allen}, the first step is to include the DW 
term in the pseudopotential form factor (Eq.~\ref{dw}) and then calculate 
the electronic structure. That is, the wavefunctions and charge densities of 
valence electrons so obtained are temperature dependent and not frozen. 
Thus, the Allen theory provides a simple way to go beyond the rigid pseudo-atom 
approximation to theoretically obtain finite temperature valence charge 
densities. 

Further, in the Allen theory~\cite{Allen}, the SE corrections to the 
electron energies are calculated from the results of the above step. 
That is, there is no further correction of the wavefunctions. 
Hence,  the  first (DW) step of the Allen theory is sufficient to obtain 
finite temperature valence charge densities. 

Figure~\ref{c-si-cden} plots the (pseudo) valence charge densities of 
(a) diamond (C$_8$) and (b) silicon (Si$_8$) at 0 K, 300 K and 600 K. 
It shows that the charge distribution in the bonding region is enhanced 
with an increase in temperature.
In general, the charge delocalization in silicon increases with increasing 
temperature.
It is interesting to note that the differences in the charge density 
distribution at 300 K and 600 K compared to that at 0 K are more 
for silicon than diamond. This is due to the much higher mean square 
displacements of silicon compared to diamond (Table ~\ref{msdv}). 
In silicon, the charge density distribution at finite temperatures shows more accumulation 
of charges in the bonding region with a reduction in the charge density in the 
vicinity of the core region. With better exchange correlation
functionals, which reduce the underestimation of the band gaps, a more 
accurate charge density distribution would be obtained.
In diamond, a similar effect is observed with a minor gain in the charge 
densities in the covalent bonding region. 

As is well known~\cite{LuZunger, YinCohen}, the pseudopotential method 
only gives pseudo-valence charge densities. They are incorrect near the 
nuclei.  However, away from the nuclei and in the valence region, the 
pseudo-valence charge densities are similar to the true valence charge 
densities~\cite{LuZunger, YinCohen}. The finite temperature charge density information in the valence region 
is valuable as this is the region where bonding effects are strongly 
manifested. The pseudo-valence charge densities plotted in Fig.~\ref{c-si-cden} thus
go beyond the rigid pseudo-atom approximation. 
Therefore, if the experimental data can also be analyzed appropriately to 
obtain finite temperature experimental charge densities, they can be compared 
with the theoretical results obtained from the Allen theory, a comparison 
that goes beyond the rigid pseudo-atom approximation.

\section{Conclusions}\lbl{concl}
In this paper, we report the implementation of the Allen theory 
with \textit{ab initio} pseudopotentials in density functional total energy 
calculations to obtain the finite temperature thermally-averaged electronic structure without 
any additional increase in the computational complexity and cost. Our results 
on diamond and silicon show that both the direct and indirect band gaps 
exhibit the ``Varshni effect'' with temperature only when the \textit{ab initio} 
pseudopotentials satisfy an additional criterion, that 
of temperature transferability. The temperature
transferability criterion is satisfied only in a small subset of the 
parameter space of static lattice in Troullier-Martins pseudopotentials 
for diamond and silicon. In these materials, the temperature transferability
is strongly affected by the choice of the cut-off radius and 
inclusion/exclusion of unbound 3d$^0$ state in the pseudopotential generation 
configuration. The finite temperature indirect band gaps in diamond and 
silicon are seen to be highly sensitive to the choice of cut-off radius.

Our results on finite temperature electronic structure show 
that the thermal vibrations affect the electron energies throughout the 
valence and conduction bands. 
We have calculated the zero-point and higher temperature band gap shifts 
in diamond and silicon for temperatures up to 1000 K. 
We compared our results on direct band gap shifts with those obtained using the 
Allen-Heine theory for the contribution from the Debye-Waller term.
For diamond, our zero-point shifts in the direct band gaps are higher by 26 
meV for the same \textit{ab initio} pseudopotentials. 
The finite temperature direct band gap shifts also show noticeable discrepancies. 
For silicon, we see a good agreement in the band gap shifts with the 
only results reported in literature (for the DW contribution) using empirical 
pseudopotentials. 
Further studies using the same temperature transferable \textit{ab initio} 
pseudopotentials and mean square displacements are essential to 
understand the discrepancies in the Allen and Allen-Heine theory band gap 
shifts, especially since the former, besides being more accurate, has a 
simpler numerical implementation.

The inclusion of Debye-Waller correction using the Allen theory provides 
a simple and theoretically justified formalism to obtain finite temperature 
valence electron charge densities.
The finite temperature charge densities obtained using Allen theory go
beyond the rigid pseudo-atom approximation, a limitation of the present
methods used in charge density studies. 

\section{Acknowledgement}\lbl{ack}

V.S. acknowledges funding support under the DST Nanomission, DST PURSE, 
DRDP and BCUD research grant from Savitribai Phule Pune University. 
B.S. and R.R.  acknowledge helpful discussions with Aditya Vishwakarma and 
Harshit Bharti on the use of OPIUM software. 

\normalem
\bibliographystyle{apsrev}
\bibliography{refs_newprb}

\end{document}